\documentclass[a4paper,11pt]{article}

\pdfoutput=1

\usepackage{amssymb,amsmath,bm}
\usepackage{a4wide}
\usepackage{color}
\usepackage{slashed}
\usepackage{cite}
\usepackage{graphicx}
\usepackage[latin1]{inputenc}
\usepackage{amsfonts}
\usepackage{lscape}
\usepackage{amsthm}
\usepackage{booktabs}
\usepackage{array}
\usepackage{rotating}
\usepackage{float}
\usepackage{multirow}

\usepackage[small,bf]{caption}
\newcommand{ \mysmall}[1]{\scriptscriptstyle #1} 

\newcommand{\GeV}{{\rm GeV}}

\newcommand{\eps}{\epsilon}
\newcommand{\ord}[1]{\mathcal{O}\left( #1 \right)}

\newcommand{\be}{\begin{equation}}
\newcommand{\ee}{\end{equation}}

\newcommand{\yu}{y_U}
\newcommand{\yd}{y_D}
\newcommand{\ye}{y_E}
\newcommand{\lu}{\lambda_U}
\newcommand{\ld}{\lambda_D}
\newcommand{\lee}{\lambda_E}

\newcommand{\yed}{y_E^\dagger}

\newcommand{\led}{\lambda_E^\dagger}

\newcommand{\eq}[1]{\begin{align}#1\end{align}}
\newcommand{\bea}{\begin{eqnarray}}
\newcommand{\eea}{\end{eqnarray}}

\newcommand{\Tr}{{\rm Tr} }

\selectfont

\begin{document}

\vspace{1cm}
\begin{titlepage}
\vspace*{-1.0truecm}
\begin{flushright}
ULB-TH/14-12 
\end{flushright}

\vspace{0.8truecm}

\begin{center}
\boldmath

{\Large\textbf{Lepton Flavor Violation in Flavored Gauge Mediation}}

\unboldmath
\end{center}

\vspace{0.4truecm}

\begin{center}
{\bf Lorenzo Calibbi$^a$, Paride Paradisi$^{b,c,d}$, Robert Ziegler$^{e,f}$}
\vspace{0.4truecm}

{\footnotesize
$^a${\sl Service de Physique Th\'eorique, Universit\'e Libre de Bruxelles, B-1050 Brussels, Belgium}\\
$^b${\sl Dipartimento di Fisica e Astronomia, Universit\'a di Padova, Via Marzolo 8, I-35131 Padova, Italy}\\
$^c${\sl INFN Sezione di Padova, Via Marzolo 8, I-35131 Padova, Italy}\\
$^d${\sl SISSA, Via Bonomea 265, I-34136 Trieste, Italy}\\
$^e${\sl Sorbonne Universit\'es, UPMC Univ Paris 06, UMR 7589, LPTHE, F-75005, Paris, France}\\
$^f${\sl CNRS, UMR 7589, LPTHE, F-75005, Paris, France}
}
\end{center}

\begin{abstract}
\noindent
We study the anatomy and phenomenology of Lepton Flavor Violation (LFV) in the context of Flavored Gauge Mediation (FGM).
Within FGM, the messenger sector couples directly to the MSSM matter fields with couplings controlled by the same dynamics 
that explains the hierarchies in the SM Yukawas.
Although the pattern of flavor violation depends on the particular underlying flavor model, FGM provides a built-in 
flavor suppression similar to wave function renormalization or SUSY Partial Compositeness. Moreover, in contrast 
to these models, there is an additional suppression of left-right (LR) flavor transitions by third-generation Yukawas that in particular provides an extra protection against flavor-blind phases. We exploit the consequences of this setup for lepton flavor phenomenology, assuming that the new couplings are controlled by simple $U(1)$ flavor models that have been proposed to accommodate large neutrino mixing angles. Remarkably, it turns out that in the context of FGM these models can pass the impressive constraints from LFV processes
and leptonic EDMs even for light superpartners, therefore offering the possibility of resolving the longstanding muon $g-2$ anomaly.
\end{abstract}

\end{titlepage}


\section{Introduction}
One of the longstanding problems in particle physics is the origin of flavor hierarchies in
the Standard Model (SM). The most popular attempt to address this problem is in terms of flavor 
symmetries in which the flavor hierarchies arise from a suitable symmetry breaking pattern.
Among the numerous possibilities, the simplest models are based on a single $U(1)$ flavor 
symmetry~\cite{Froggatt:1978nt,Leurer:1992wg,Leurer:1993gy,Binetruy:1994ru}.
In the quark sector this ansatz works pretty well and can account for all hierarchies in quark
masses and mixing, with an order-of-magnitude prediction $V_{ub} \sim V_{us} V_{cb}$ that is in
good agreement with data.
Also in the lepton sector a single $U(1)$ works very well, since charge lepton mass hierarchies
can arise from large charge differences of right-handed leptons, while large mixing angles are
due to small charge differences of left-handed leptons. In this way $U(1)$ models can naturally 
realize the paradigm of an ``anarchical" structure~\cite{Hall:1999sn, Haba:2000be, Altarelli:2002sg,deGouvea:2003xe} 
in lepton mixing, which
has recently received renewed attention~\cite{deGouvea:2012ac ,Altarelli:2012ia,Bergstrom:2014owa} 
after the reactor neutrino angle $\theta_{13}$ turned out to be sizable.

Independently of the nature of the underlying flavor symmetry, the crucial question about these
kind of models regard their predictivity. Since flavor models aim at explaining the origin of 
dimensionless Yukawa couplings, there is no preferred mass scale of the new degrees of freedom.
As new effects in the SM flavor sector are suppressed by this mass scale, there are no observable 
deviations from the SM flavor predictions, unless this scale is unexpectedly light~\cite{Calibbi:2012at}. 
Therefore the only way to test these models in laboratory experiments for a high-scale flavor 
sector is the presence of new physics around the TeV scale, as suggested by the hierarchy problem. 
If such physics comes with a flavor structure, it can possibly carry down the information of the 
high-scale flavor sector to the TeV scale and lead to testable predictions for precision flavor 
observables. The prime example is Supersymmetry (SUSY), which in the case of high-scale mediation 
of SUSY breaking around or above the flavor sector scale directly contains the imprint of the flavor 
symmetry in the soft-breaking sfermion masses.

However, within the context of Gravity Mediation simple $U(1)$ 
models are in big trouble as the suppression of flavor violation is too weak. The reason is that 
off-diagonal entries in the left-handed and right-handed sfermion mass matrices are suppressed by the differences of the corresponding 
charges due to their non-holomorphic nature. In the left-handed sector these charge differences are directly 
related to mixing angles, which for the first two generations are sizable both in the lepton and quark sector. 
Since the strongest constraint precisely arise from observables involving light families, like $\eps_K$ 
and $\mu \to e \gamma$, such $U(1)$ flavor models in the context of Gravity Mediation are essentially 
incompatible with SUSY around the TeV scale.\footnote{This conclusion holds even under the assumption 
of a mechanism inducing degenerate sfermion masses at the SUSY breaking mediation scale. In fact, 
the assumed flavor universality is broken by the renormalization group evolution of the soft masses 
down to the flavor breaking scale, so that large flavor mixing is anyway generated at the level 
predicted by the $U(1)$ symmetry \cite{Calibbi:2012yj}.}

The situation is completely different in Gauge Mediation (see Ref.~\cite{Giudice:1998bp}), where 
the SUSY breaking and the flavor sector can be decoupled. Indeed, if the flavor scale is much higher than 
the SUSY messenger scale then soft masses are screened from the high-energy flavor sector and have a flavor 
structure determined only by SM Yukawas, thus realizing the paradigm of Minimal Flavor Violation 
(MFV)~\cite{D'Ambrosio:2002ex}. While this scenario provides a very appealing mechanism to solve the SUSY 
flavor problem, the imprint of the flavor sector in low-energy physics and thus the possibility to test 
flavor symmetry models is completely lost.
  
It is therefore interesting to construct extensions of Minimal Gauge Mediation (MGM) that re-introduce 
the dependence on the underlying flavor sector, and thus lead to a broad variety of sfermion flavor 
structures beyond MFV. An example for such extensions is provided by a class of models that has been 
dubbed ``Flavored Gauge Mediation" (FGM)~\cite{Shadmi:2011hs}. In these scenarios, new direct couplings 
between the messengers and the MSSM matter fields are introduced with a flavor structure that is assumed to be 
controlled by the same underlying flavor symmetry that explains the smallness of the Yukawas.\footnote{Such 
matter-messenger couplings have recently received new interest, as they allow to obtain a large Higgs mass 
with light stops by generating non-vanishing A-terms at the messenger scale~\cite{Evans:2011bea, Draper:2011aa, Evans:2012hg, Kang:2012ra, Craig:2012xp, Albaid:2012qk, Abdullah:2012tq, Byakti:2013ti, Evans:2013kxa, Jelinski:2013kta, Jelinski:2014uba}.}
These couplings generate new contributions to sfermion masses (on top of of the flavor-universal MGM ones) that are controlled by the underlying flavor symmetry. Interestingly, due to the loop origin of the soft terms, there is a built-in suppression of flavor violation that is independent of the underlying flavor model~\cite{Calibbi:2013mka}. This implies that even single $U(1)$ flavor models are perfectly viable (in contrast to Gravity Mediation), as the flavor pattern of the resulting sfermion masses resembles the suppression in wave function renormalization~\cite{Isidori,Pokorski} or SUSY Partial Compositeness~\cite{Nomura,Lodone}. Moreover, in contrast to those scenarios, there is also a built-in suppression of LR flavor transitions and in particular flavor-blind phases by third-generation Yukawas, which becomes very efficient in the down and charged-lepton sector provided $\tan \beta$ is not very large.

While in Ref.~\cite{Calibbi:2013mka} we have focused on the quark sector, in this paper we
analyze the impact of FGM models with underlying $U(1)$ flavor models on the lepton sector.
There are good arguments that motivate this study:
i) in contrast to the quark sector the large neutrino mixing angles require milder hierarchies in left-handed charges, leading 
in turn to weaker suppression in the left-handed slepton sector and therefore potentially large
effects in LFV processes, ii) the experimental bounds on
LFV channels with an underlying $\mu \to e$ transition as well as the electron EDM underwent
recently a very significant improvement challenging many models with New Physics (NP) at the TeV scale, even with modest
sources of flavor violation.
Therefore the major aim of this work is to analyze whether and to which 
extent single $U(1)$ flavor models for the lepton sector are viable in the context of FGM. 
A related question is whether we can account for the current muon $g-2$ anomaly, that is if 
light sleptons are still allowed by the LFV and EDM bounds (for a general discussion on the interrelationship of leptonic dipoles see Ref.~\cite{Giudice:2012ms}).

The rest of the paper is organized as follows: in Section 2 we recall the main ingredients
of FGM models providing explicit expressions for the soft masses in the slepton sector.
Concrete examples of $U(1)$ leptonic flavor models and their imprint in the soft sector
are presented in Section 3. The low-energy phenomenology of FGM models supplemented by
the above $U(1)$ flavor models is studied in Section 4. In Section 5, we compare
the flavor structure of the soft terms and related phenomenological implications of FGM
models to $U(1)$ models with Gravity Mediation and models with SUSY Partial Compositeness.
We conclude in Section 6. In an appendix we collect the formulae for LFV branching ratios, lepton anomalous magnetic moments and lepton electric dipole moments using a generalized mass insertion approximation without assuming large $\tan \beta$, thus improving on existing results that take into account only the $\tan \beta$ enhanced terms.

\section{Flavored Gauge Mediation}

We begin with a brief review of Minimal Gauge Mediation (see Ref.~\cite{Giudice:1998bp}). In this scenario
$N$ copies of heavy chiral superfields $\Phi_i + \overline{\Phi}_i$ in ${\bf 5+\overline{5}}$
of SU(5) are introduced. These messenger fields couple directly to the SUSY breaking sector,
which is effectively parameterized by a single spurion field $X$ that gets a vev
$\langle X \rangle = M + F \theta^2$. Through the following superpotential coupling
\be
\label{WMGM}
W \supset  X  \overline{\Phi}_i \Phi_i, \qquad i=1 \hdots N\,,
\ee
the messengers acquire large supersymmetric mass terms $M$ and SUSY breaking masses proportional to $F$. By integrating out the messengers 
at loop-level, soft terms are generated. At the messenger scale, A-terms vanish and gaugino masses and sfermion masses are given by
\begin{gather}
 \label{eq:GMSB-Mi}
 M_i(M)  = N \frac{\alpha_i(M)}{4 \pi} ~\Lambda,  \qquad \qquad \Lambda =  \frac{F}{M}, \\
 \label{eq:GMSB-soft}
 m^2_{\tilde f}(M)  = 2 N \sum_{i=1}^3 C_i(f)~ \frac{\alpha^2_i(M)}{(4 \pi)^2} ~\Lambda^2, \qquad 	f=q,\,u,\,d, \ldots,
\end{gather}
where $C_i(f)$, $i=1,2,3$ is the quadratic Casimir of the representation of the field
$f$ under the gauge group ${\rm SU(3)}\times {\rm SU(2)}\times {\rm U(1)}$.

Since the messengers have the same gauge quantum numbers as the MSSM Higgs fields,
in addition to the Yukawa couplings
\be
 W \supset  (\yu)_{ij} Q_i U_j H_u +  (\yd)_{ij} Q_i  D_j H_d +  (\ye)_{ij} L_i  E_j H_d\,,
\label{eq:W}
\ee
also direct couplings of messengers to MSSM fields are allowed by the gauge symmetries.
If we restrict to R-parity even messenger fields,\footnote{For R-parity odd messengers similar couplings to Higgs fields are allowed and have been discussed in the literature, see e.g.~\cite{Evans:2013kxa}. Note that in this way one preserves the MFV structure of MGM. Here we are instead interested in non-trivial flavor structures. } the messengers can couple only to the
MSSM matter fields. For the messenger doublets these couplings read in general
\begin{align}
\label{newcoupl}
\Delta W & = (\lu)_{ij} Q_i U_j \Phi_{H_u} +  (\ld)_{ij} Q_i  D_j \overline{\Phi}_{H_d} +  (\lee)_{ij} L_i  E_j \overline{\Phi}_{H_d},
\end{align}
where $\Phi_{H_u}, \overline{\Phi}_{H_d}$ denote the SU(2) doublet
components of the $\bf 5, \bf \overline{5}$ messengers, and we restricted to the case of one messenger pair for simplicity.

The presence of direct messenger-matter couplings gives rise to new contributions to sfermion masses 
and A-terms with a flavor structure that depends on the new parameters $\lambda_{ij}$. If these couplings 
were flavor-anarchic $\ord{1}$ numbers, the elegant solution of Gauge Mediation to the SUSY flavor problem 
would be completely spoiled. Therefore it is usually assumed that all direct couplings of the messengers 
to matter fields vanish, which can be enforced for example by introducing a new $Z_2$ symmetry under which 
MSSM fields are even and messengers are odd. Note that this symmetry extends to a full accidental 
$U(1)$ symmetry in the case of one messenger pair 
\eq{ \label{U1M}\Phi & \to e^{i \alpha} \Phi, &  \overline{\Phi} & \to e^{- i \alpha} \overline{\Phi}. }
However, in order to preserve the neat solution of the SUSY flavor problem in MGM, it is enough that the new couplings in Eq.~(\ref{newcoupl}) are just sufficiently small. Such small couplings can be easily motivated in the context of flavor models, since they break the global flavor symmetries of MSSM kinetic terms exactly as the Yukawas, and therefore they can naturally have a similar hierarchical structure. This can be realized in explicit flavor models in which the messenger fields transform like the Higgs fields (in particular one can choose that they do not transform at all under the flavor sector), which implies that the new couplings have the same parametric suppression as the Yukawas, 
\begin{align}
\lambda_U & \sim y_U, & \lambda_D & \sim y_D, & \lambda_E & \sim y_E.
\end{align}
Following Ref.~\cite{Shadmi:2011hs}, we refer to these kind of models as ``Flavored Gauge Mediation" (FGM). 

The new contributions to soft terms induced by the couplings in Eq.~(\ref{newcoupl}) can be calculated using the general expressions in Ref.~\cite{Evans:2013kxa}. At leading order in SUSY breaking one finds new contributions to sfermion masses at 2-loop and non-vanishing A-terms at 1-loop. While these new effects can have interesting consequences for the low-energy spectrum~\cite{Shadmi2,Galon:2013jba}, here we are mainly interested in the flavor structure of the new contributions to sfermion masses, in particular in the slepton sector. Therefore we will now take a bottom-up point of view and restrict the analysis to the consequences of the presence of the $\lambda_E$ coupling for the lepton sector. We will not discuss the impact of other possible messenger-matter couplings on the low-energy spectrum, in particular the mass of the lightest Higgs boson. We just note that the Higgs mass does not represent a serious constraint in these kind of models, and can be due to large A-terms or an implementation in the NMSSM. The latter also represents a natural possibility to generate the $\mu-$term and to elegantly solve the $\mu-B_\mu$ problem of Gauge Mediation, since in the NMSSM the general structure of FGM motivates a direct coupling of the NMSSM singlet to the messengers which can easily allow for correct EWSB~\cite{Craig:2012xp, new}. 

Furthermore, let us notice that the couplings $\lambda_E$ do not deform the spectrum predicted by the underlying gauge mediation
scheme, at least for low to moderate values of $\tan\beta$, as we are going to consider in the next sections. In particular, if $m_h\approx 126$ GeV is accounted for by a large top A-term, induced by an $\mathcal{O}(1)$ coupling $(\lu)_{33}$
in Eq.~(\ref{newcoupl}), the spectrum would resemble the one discussed e.g.~in Ref.~\cite{Calibbi:2013mka}.
This would have interesting consequences for the leptonic sector we consider here, since 
$(\lu)_{33} = \mathcal{O}(1)$ also suppresses the masses of the left-handed sleptons, through an induced Fayet-Iliopoulos term,
thus naturally accomodating the Higgs mass with a light slepton spectrum that can give a sizable contribution
to the muon $g-2$ \cite{Evans:2012hg,Calibbi:2013mka}. 

For soft terms in the slepton sector we use the conventions
\begin{align}
{\cal L} & \supset - \left( (\tilde{m}^2_L)_{ij} L_i L_j^\dagger + (\tilde{m}^2_E)_{ij} E_i^\dagger E_j  +  (A_e)_{ij} L_i E_j  H_d \right)|_{scalar} \nonumber \\
& =   - \left( \tilde{l}_L^T \tilde{m}^2_L  \tilde{l}_L^* + \tilde{e}_R^T (\tilde{m}^2_E)_{ij}  \tilde{e}_R^* +  \tilde{l}_L^T A_e \tilde{e}_R^* H_d \right),
\end{align}
where the first line denotes the scalar components of superfields. Using the results of Ref.~\cite{Evans:2013kxa}, the presence of $\lambda_E$ gives rise to the following expressions for the non-holomorphic masses\footnote{We only consider terms in leading order in $\Lambda^2/M^2$, i.e.~a  messenger scale that is not particularly low.}
\bea
{\tilde m}^2_{L} & = & \frac{\Lambda^2}{256 \pi^4}
\bigg[ 
N \left(\frac{3}{2} g_2^4 + \frac{3}{10} g_1^4 \right)
-\left(\frac{9}{5}g_1^2 + 3 g_2^2 \right)\lee\led 
\nonumber\\
&+& 3 \lee\led\lee\led + 2\lee\yed\ye\led -2\ye\led\lee\yed
\nonumber\\
&+& \lee\led \Tr\left(\lee\led \right) +
\ye \led \Tr\left( \lee \yed \right) + \lee \yed \Tr\left( \ye \led \right)
\bigg]\,,
\label{eq:m2L}
\eea
and
\bea
{\tilde m}^2_E & = & \frac{\Lambda^2}{256 \pi^4} 
\bigg[ 
\frac{6}{5}g_1^4 N
-\left(\frac{18}{5}g_1^2 + 6 g_2^2  \right)\led\lee + 6 \led\lee\led\lee
\nonumber\\
&+& 2 \led\ye\yed\lee - 2 \yed\lee\led\ye + 2\led\lee\Tr\left(\lee\led \right)
\nonumber\\
&+& 2 \led\ye\Tr\left(\lee\yed\right) + 2 \yed\lee\Tr\left(\ye\led\right)
\bigg]\,,
\label{eq:m2e}
\eea
while the A-terms are given by
\bea
A_E = -\frac{\Lambda}{16 \pi^2} \left(\lee \led \ye + 2\ye  \led \lee \right)\,.
\label{eq:Ae}
\eea
Note that the flavor dependence of the above expressions can be obtained using a simple spurion analysis, 
taking into account also the $U(1)_M$ ``messenger number" in Eq.~(\ref{U1M}) as a spurious
symmetry under which the new couplings are charged. The $U(1)_M$ symmetry prevents
terms like $\lee\yed$ and $\led\ye$ for the non-holomorphic masses ${\tilde m}^2_{L}$ and
${\tilde m}^2_{E}$, respectively, and terms like $\lee$, $\lee\yed\lee$ and $\led\ye\led$ for 
the A-terms. As a result, the A-terms are partially aligned to the Yukawa couplings and their
diagonal components are necessarily real and therefore do not induce contributions to the EDMs.

For future convenience, we define the flavor violating mass insertions (MIs) as usual 
\begin{align}
(\delta_{LL}^e)_{ij} & = \frac{({\tilde m}^2_{L})_{ij}}{{\tilde m}^2_{L}}~, &
(\delta_{RR}^e)_{ij} & = \frac{({\tilde m}^2_{E})_{ij}}{{\tilde m}^2_{E}}~, &
(\delta_{LR}^e)_{ij} & = \frac{v_d(A_{E})_{ij}}{{\tilde m}_{L}{\tilde m}_{E}}~. 
\end{align}
In the limit of $\ye,\lee \ll 1$, i.e.~for moderate/low $\tan\beta$ values, 
we obtain the following approximate expressions at the messenger scale: 
%
%
\begin{align}
(\delta_{LL}^e)_{ij} & \simeq - \left(\frac{10 g_2^2 + 6 g_1^2}{5 g_2^4 + g_1^4}\right) \frac{(\lee\led)_{ij}}{N}\,, &
(\delta_{RR}^e)_{ij} & \simeq - \left(\frac{5 g_2^2 + 3 g_1^2}{g_1^4}\right) 
\frac{(\led\lee)_{ij}}{N}\,,
\label{dLL_dRR}
\end{align}
\begin{align}
(\delta_{LR}^e)_{ij} \simeq - \frac{1} {\sqrt{\frac{3N}{\sqrt{5}}} g_1g_2}~
\frac{m^e_j (\lee\led)_{ij} + 2 m^e_i (\led\lee)_{ij}}{\sqrt{\tilde{m}_L \tilde{m}_E}}\,.
\label{dLR}
\end{align}
Few comments are in order:
\begin{itemize}
 \item The above MIs, as well as all superpotential couplings, are defined in the basis
where we define the flavor model. In order to study their phenomenological consequences,
we go to the mass basis for the charged lepton Yukawas by means of the  rotation
$y_E \to V_{EL}^T y_E V_{ER} = y_E^{diag}$. Under this change of basis, the spurion
$\lambda_E$ transforms accordingly. However, one can easily check that in $U(1)$ models
with non-negative charges the parametric flavor suppression remains the same and only
the ${\cal O}(1)$ coefficients change. We therefore simply ignore these differences,
that is we take $V_{EL}^T \lambda_E V_{ER} \sim \lambda_E$.

\item Interestingly, the diagonal A-terms are real. As a result, the leading CP violating
phases generating the EDMs can only arise at higher order in the MIs, through the combination 
$(\delta_{LL}^e)_{ik} (\delta_{LR}^e)_{kj}$, $(\delta_{LR}^e)_{ik} (\delta_{RR}^e)_{kj}$,
and $(\delta_{LL}^e)_{ik} (\delta_{LR}^e)_{kk} (\delta_{RR}^e)_{kj}$ when $ij=11$. 
This however leads to an additional suppression by powers of 
$(\lambda_E)_{33} \sim y_\tau$.\footnote{Notice that the $\mu$- and $B_\mu$-terms, which are not
controlled by GMSB, could still introduce CP violating phases depending on the underlying mechanism 
that generates them. However, if this mechanism is such that the phases of $\mu$ and $B_\mu$ are 
correlated to the phase of $\Lambda$, then no phases arise from this sector~\cite{Giudice:1998bp}. }
\item As a consequence of Eqs.~(\ref{dLL_dRR}),(\ref{dLR}), the naive expectations for the MIs are
enhanced, for a given number of messengers $N$, by large (mediation scale dependent) gauge
factors. This is especially true in the case of $(\delta_{RR}^e)_{ij}$ and, to less
extent, also in the cases of $(\delta_{LL}^e)_{ij}$ and $(\delta_{LR}^e)_{ij}$.
\end{itemize}
In the following, we will analyze the impact of our FGM model on the branching ratio of
$\mu\to e \gamma$ and the electron EDM which are the most powerful probes of new physics
in the leptonic sector. To do so, we need to specify the underlying flavor model that 
controls the flavor structure of the new couplings.

\section{Flavored Gauge Mediation and $U(1)$ Flavor Models}

While the results of the last section can be applied to any flavor model that predicts the flavor structure of $y_E$ and therefore $\lambda_E$, 
in this section we concentrate on simple $U(1)$ flavor models. We first recall the basic structure of these models, then we analyze their 
predictions for the soft terms in the lepton sector in the context of our FGM model.

\subsection{$U(1)$ Flavor Models}

In the simplest realization of these models the flavor symmetry is spontaneously broken by the vev of a single ``flavon" field with negative unit charge. Yukawa couplings then arise from higher-dimensional operators that involve suitable powers of the flavon to make the operator invariant under the $U(1)$ symmetry, with some undetermined coefficients that are assumed to be ${\cal O}(1)$. The suppression scale is the typical scale of the flavor sector that could correspond to the mass scale of Froggatt-Nielsen messengers in explicit UV completions. The Yukawas then depend only on powers of the ratio $\eps$
of flavon vev and flavor scale, which typically is taken to be of the order of the Cabibbo angle $\eps \sim 0.2$. If we restrict to models where only the matter fields are charged, i.e.~$H_u = H_d = 0$, we get for the lepton Yukawa couplings
\eq{(y_E)_{ij} \sim \eps^{L_i + E_j},}
where $L_i$ and $E_i$ stand for the $U(1)$ charges of the left-handed and right-handed leptons, respectively.
The neutrino sector depends on the origin of neutrino masses. If neutrinos are Dirac, then the
Yukawa coupling takes the same form as the charged lepton Yukawa above with $E_j \to N_j$. In this case, 
the left-handed rotations $V_{EL}, V_{NL}$ for the charged lepton and neutrino sectors, respectively, 
and therefore the PMNS matrix $V_{PMNS}$, have the same parametric structure
\eq{ \label{PMNS} (V_{PMNS})_{ij} \sim (V_{EL})_{ij} \sim (V_{NL})_{ij} \sim \eps^{|L_i - L_j|}.}
Large neutrino mixing angles can therefore be reproduced by taking small left-handed charge differences $L_i-L_j$. Instead small neutrino masses can be accommodated by taking sufficiently large charges $N_i$ of right-handed neutrinos. A more plausible explanation of light neutrinos can be achieved if they originate from the Weinberg operator
\eq{ \Delta W = \frac{(y_{ll})_{ij}}{\Lambda} L_i L_j H_u H_u,}
with a flavor structure determined by the $U(1)$ symmetry
\eq{(y_{ll})_{ij} \sim \eps^{L_i + L_j}.}
In this way the smallness of neutrino masses can be elegantly explained by assuming a large UV scale $v_u/\Lambda \ll 1$, 
but the prediction for the parametric structure of the left-handed neutrino rotations and therefore for the PMNS matrix 
does not change, and we still get the result of Eq.~(\ref{PMNS}).
%
%
One possibility for an explicit UV completion is the type-I seesaw mechanism.
In this scenario, one adds three heavy right-handed neutrinos and Dirac Yukawa couplings
\eq{ \Delta W = (y_\nu)_{ij} L_i N_j H_u + \frac{1}{2} (M_N)_{ij} N_i N_j,}
with their flavor structure given by 
\eq{(y_\nu)_{ij} & \sim \eps^{L_i + N_j} & (M_N)_{ij} & \sim M_N \eps^{N_i + N_j}.}
Integrating out the right-handed neutrinos generates the Weinberg operator with a 
coefficient given by
\eq{ \frac{(y_{ll})_{ij}}{\Lambda} = - \frac{1}{2} (y_\nu M_N^{-1} y_\nu^T)_{ij}.} 
Note that in the simple $U(1)$ models that we will consider here, the parametric flavor structure
of the coefficient of the Weinberg operator is the same as in the effective theory
\eq{(y_{ll})_{ij} \sim \eps^{L_i + L_j},}
and therefore we recover the same estimate for the PMNS matrix as in Eq.~(\ref{PMNS}). 

Various $U(1)$ models have been discussed in the literature, see e.g.~\cite{Haba:2000be, Altarelli:2004za, Buchmuller:2011tm, Altarelli:2012ia}. There is some ambiguity in the choice of charge assignments, since
$\epsilon$ is typically not a very small parameter (one has $\epsilon \approx 0.2 \div 0.5$) so that the unknown ${\cal O}(1)$ 
parameters can account for one or two units of charge differences. Here we choose to consider just two representative models that 
have been presented in Ref.~\cite{Altarelli:2012ia} and more carefully analyzed in Ref.~\cite{Bergstrom:2014owa}. The first one, 
``Anarchy", features degenerate charges of left-handed lepton doublets, so that all mixing angles are predicted to be ${\cal O}(1)$. 
The second one, ``Hierarchy", has non-degenerate charges in order to account for the relative smallness of $\theta_{13}$ and $\Delta m^2_{solar}/\Delta m^2_{atm}$. 
Other models that have been considered in Ref.~\cite{Buchmuller:2011tm, Altarelli:2012ia} fall in between these two models for what regards their phenomenological 
consequences in FGM. The charge assignments of the two models are given by 
\begin{itemize}
\item Anarchy 
 \begin{align} E_i & = (3,2,0) & L_i & = (L_3,L_3,L_3),& N_i & = (0,0,0), & \epsilon_A \approx 0.2\,. \end{align}
 \item Hierarchy 
  \begin{align} E_i & = (5,3,0) & L_i & = (2+L_3,1+L_3,L_3),& N_i & = (2,1,0), & \epsilon_H \approx 0.3\,. \end{align}
\end{itemize}
For simplicity the expansion parameters are taken here as the central values of the accurate fit in Ref.~\cite{Bergstrom:2014owa},
although there is of course some range due to the unknown order one coefficients.  We will use these values in the numerical 
analysis of Section 4. Note the dependence on an overall charge shift $L_3$ that essentially corresponds to $\tan \beta$.

\subsection{Application to FGM}

We now discuss the implementation of the above $U(1)$ models in FGM. For this we only have to
specify the charges of the messengers. While in principle they can be arbitrary, we only consider
the simplest choice in which they have the same charges as the Higgs fields, i.e.~they transform 
trivially under the flavor symmetry $\Phi = \overline{\Phi} = 0$. This immediately implies that
the new couplings of matter fields to the messengers have exactly the same parametric suppression
as the corresponding matter-Higgs couplings, but with different ${\cal O}(1)$ coefficients.
In order to see this point more explicitly, we go to the mass basis for the charged leptons
by means of the superfield transformations
\begin{align}
L & = \begin{pmatrix} L_N \\ 
L_E \end{pmatrix} \to  \begin{pmatrix} V_{NL} L_N \\ 
V_{EL} L_E \end{pmatrix}, & E & \to V_{ER} E,
\end{align}
so that
\begin{align}
y_E & \to V_{EL}^T y_E V_{ER} = y_E^{diag}.
\end{align}
Note that in $U(1)$ models the rotations have the simple parametric structure
\eq{ (V_{EL})_{ij} & \sim \eps^{|L_i - L_j|} & (V_{ER})_{ij} & \sim \eps^{|E_i - E_j|}.}
The spurion $\lambda_E$ transforms accordingly under the above rotations. However, it is
straightforward to check that its parametric flavor suppression remains unchanged and only
the ${\cal O}(1)$ coefficients do change so that $V_{EL}^T \lambda_E V_{ER} \sim\lambda_E$.
As a result, since the flavor suppression of $\lee$ is the same of $\ye$, one gets in the mass basis
\begin{equation}
\label{lErel}
(\ye)_{ii} = a_{ii} \epsilon^{L_i + E_i}\,,
\qquad
(\lee)_{ij} = \kappa_{ij} \epsilon^{L_i + E_j}\,,
\end{equation}
where $a_{ii}$ and $\kappa_{ij}$ account for unknown, flavor dependent, ${\cal O}(1)$ coefficients.
Assuming hierarchical charges ($E_3 \le E_{k}$ etc.), we finally find the following MIs:
\begin{align}
(\delta_{LL}^e)_{ij}  & \sim  \epsilon^{L_i + L_j + 2 E_3} \sim y^2_\tau \epsilon^{L_i + L_j - 2 L_3},
\label{lleLLsupp}
\\
(\delta_{RR}^e)_{ij}  & \sim  \epsilon^{E_i + E_j + 2 L_3} \sim y^2_\tau \epsilon^{E_i + E_j - 2 E_3},
\label{lleRRsupp}
\end{align}
and similarly 
\begin{align}
(\delta_{LR}^e)_{ij} \sim y^2_\tau~\frac{m^e_j~\epsilon^{L_i + L_j - 2 L_3} +
2 m^e_i~\epsilon^{E_i + E_j - 2 E_3}}{\sqrt{\tilde{m}_L \tilde{m}_E}},
\end{align}
where the overall coefficients are given by a calculable part that can be read off from 
Eqs.~(\ref{dLL_dRR}),(\ref{dLR}) and an unknown ${\cal O}(1)$ coefficient coming from Eq.~(\ref{lErel}).

As already discussed, since the diagonal A-terms are real, the EDMs can be only generated
by means of the combination of MIs $(\delta_{LL}^e)_{ik} (\delta_{LR}^e)_{kj}$, $(\delta_{LR}^e)_{ik} (\delta_{RR}^e)_{kj}$, and $(\delta_{LL}^e)_{ik} (\delta_{LR}^e)_{kk} (\delta_{RR}^e)_{kj}$
when $ij=11$. In particular, it turns out that the leading effect is captured by
\begin{align}
(\delta_{LL}^e)_{i3} (\delta_{LR}^e)_{33} (\delta_{RR}^e)_{3j} & \sim \frac{\mu\tan\beta \, m_\tau}{\tilde{m}_L \tilde{m}_E } y_\tau^3 \epsilon^{L_i + E_j},
\label{triple_MI}
\end{align}
which involves additional powers of $y_\tau$. In principle, the effective MI of Eq.~(\ref{triple_MI}) also contributes 
to $\mu\to e\gamma$ when $ij=12,21$ however, within our models, single MI contributions always dominate.
\begin{figure}
\centering
\includegraphics[scale=0.4]{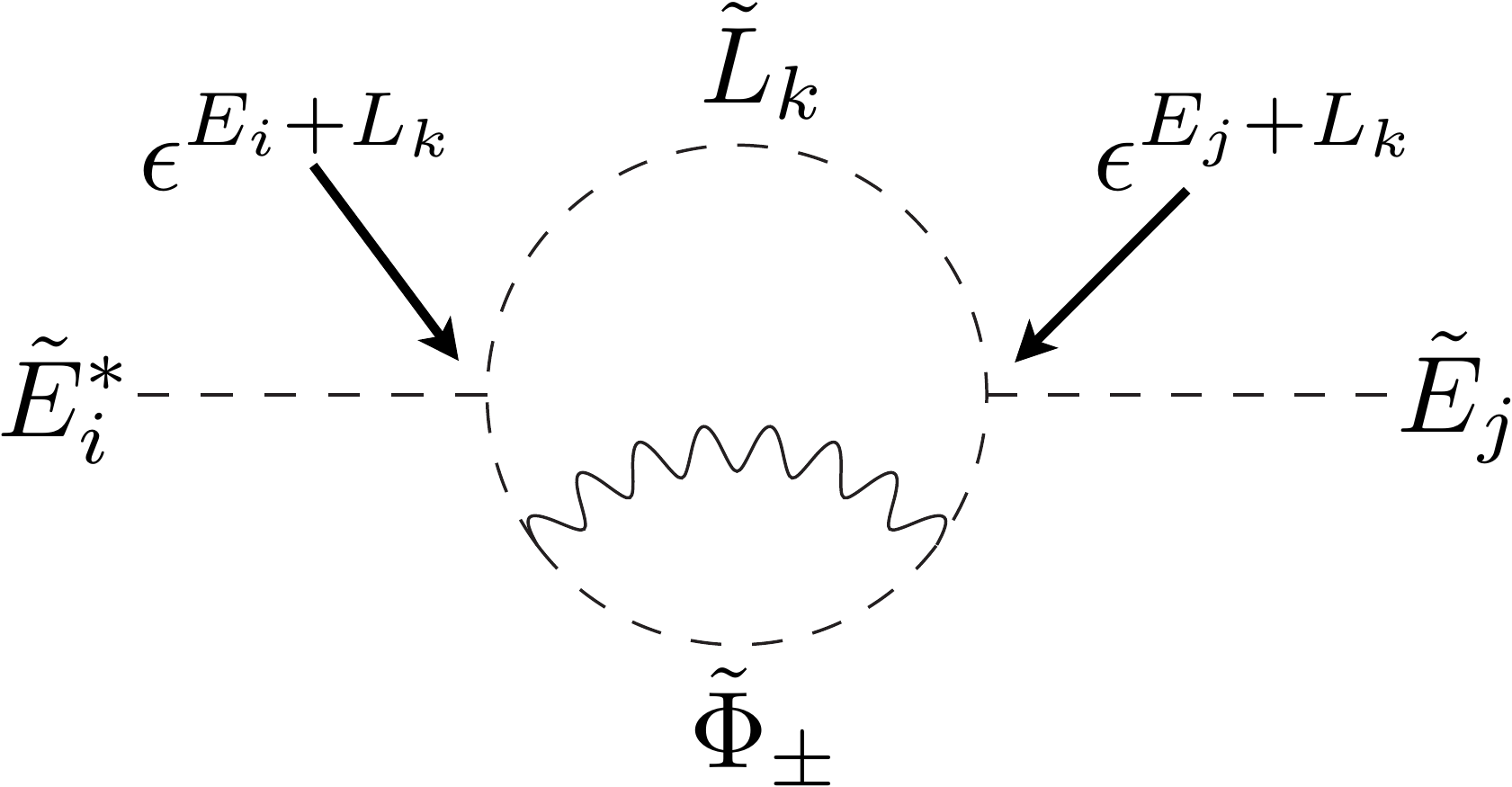}
\caption{Example diagram for the 2-loop generation of $(\delta^e_{RR})_{ij}$. $\tilde{L}_i,\tilde{E}_i$ denotes the scalar components of the superfields $L_i,E_i$ and $\tilde{\Phi}_\pm$ denotes the scalar mass eigenstates of the messengers.}
\label{diagramsfermions}
\end{figure}
\begin{figure}
\centering
\includegraphics[scale=0.4]{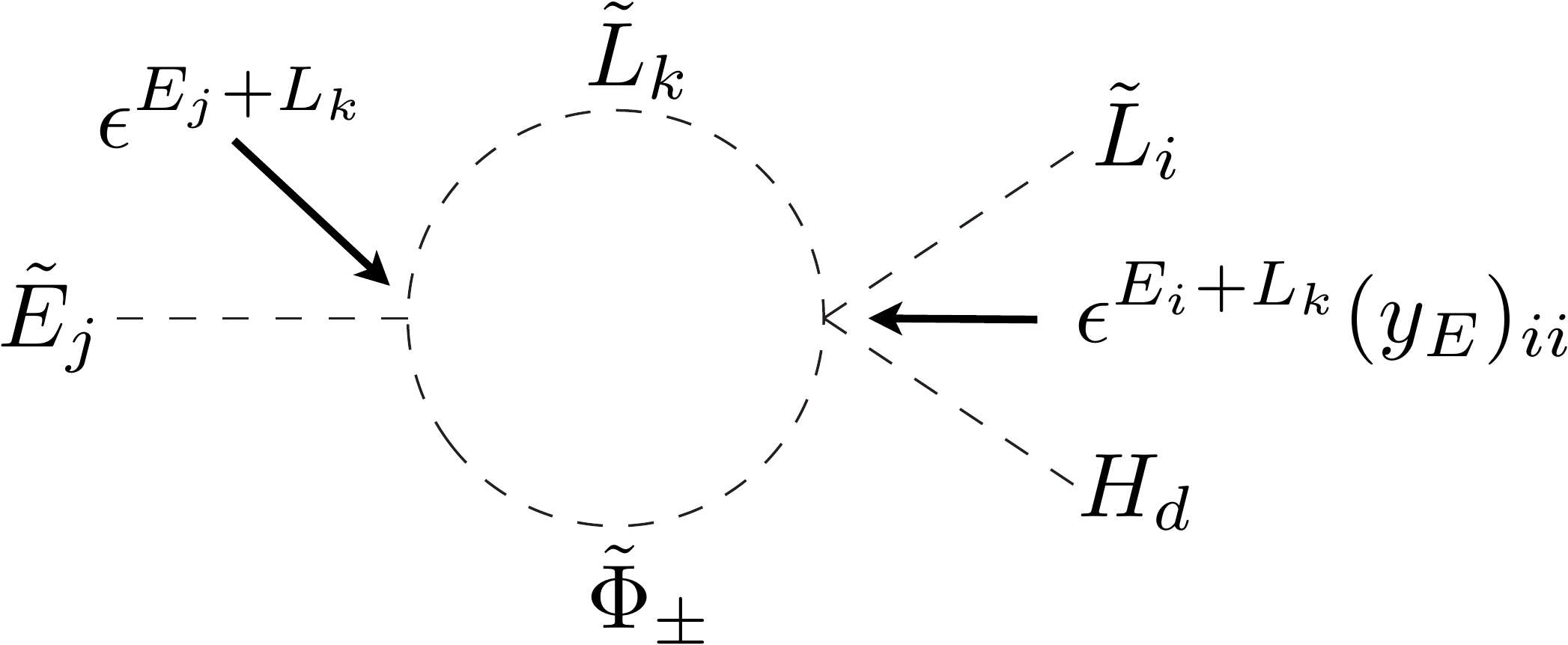}
\caption{Example diagram for the 1-loop generation of $(\delta^e_{LR})_{ij}$ in the fermion mass basis.}
\label{diagramaterms}
\end{figure}

We conclude this section with a discussion of the general structure of the flavor suppression in these terms.
First of all, note that LL and RR mass insertions are suppressed by powers of the spurion that are the {\it sum} 
of U(1) charges, in contrast to the leading order terms allowed by the symmetry that have powers given by charge 
differences. The origin of this suppression is due to the fact that the $U(1)$ controls soft terms only indirectly 
via the messenger sector, which in turn generates soft terms only at loop level, thus leading to a double suppression 
by small couplings. As can be seen e.g.~in Eq.~(\ref{lleRRsupp}) and the corresponding 2-loop diagram for 
$\delta^e_{RR}$ in Fig.~\ref{diagramsfermions}, this suppression can be split into two parts, one given by the sum 
of charges of the external sfermions and the second by (twice) the charge of the field that runs in the loop together 
with the messenger.
As we will discuss later on, the first suppression is exactly the same as in SUSY Partial Compositeness, 
while the second can lead to a further suppression by powers of $y_\tau$.

Turning to LR mass insertions, again the loop origin implies a much stronger suppression than the leading order term 
$\epsilon^{L_i + E_j}$ respecting the $U(1)$ symmetry, see Fig.~\ref{diagramaterms}. This suppression is partially 
due to the alignment to Yukawas in the pure LR term, which potentially can be avoided in the effective LR terms 
at the price of an additional suppression by powers of $y_\tau$. This is also the only way in which phases can arise 
in the diagonal elements, since the pure LR term is always the product of a hermitian and a real diagonal matrix.

\section{Flavor Phenomenology}

\begin{table}[t!]
\centering
\begin{tabular}{|c|c|c|}
\hline
LFV Process & Present Bound & Future Sensitivity  \\
\hline
$\mu \to e \gamma$ & $5.7 \times 10^{-13}$ \cite{Adam:2013mnn} & $\approx 6 \times 10^{-14}$ \cite{Baldini:2013ke}  \\
$\mu \to 3 e$ & $1.0 \times 10^{-12}$\cite{Bellgardt:1987du} & $\approx 10^{-16}$ \cite{Blondel:2013ia}\\
$\mu^-$ Au $\to$ $e^-$ Au & $7.0 \times 10^{-13}$ \cite{Bertl:2006up} & $ ? $  \\
$\mu^-$ Ti $\to$ $e^-$ Ti & $4.3 \times 10^{-12}$ \cite{Dohmen:1993mp} & $?$ \\
$\mu^-$ Al $\to$ $e^-$ Al & $-$  & $\approx 10^{-16}$ \cite{comet,mu2e} \\
\hline
Electron EDM & Present Bound & Future Sensitivity  \\
\hline
$d_e ({\rm e~cm})$ & $8.7 \times 10^{-29}$ \cite{Baron:2013eja} & $?$ \\
\hline
\end{tabular}
\caption{Current experimental bounds and future sensitivities for some low-energy LFV
         observables and the electron EDM.}
\label{tab:LFV}
\end{table}

We are now ready to discuss the lepton flavor phenomenology of the FGM model, which includes
LFV processes with an underlying $\mu\to e$ transition, the electron EDM $d_e$ and the anomalous 
magnetic moment of the muon $a_\mu\equiv(g-2)/2$. The current experimental bounds and future sensitivities for some
of the most relevant LFV channels and for $d_e$ are reported in Table~\ref{tab:LFV}.
On the other hand, $a_\mu$ currently shows a discrepancy between the SM prediction and 
the experimental value~\cite{bnl, Jegerlehner:2009ry, HLMNT11, DHMZ11}
\be
\Delta a_\mu = a_\mu^{\mysmall \rm EXP}-a_\mu^{\mysmall \rm SM} = 2.90 (90) \times 10^{-9}\,.
\ee
One of the goals of the present paper is to investigate whether it is possible to explain this anomaly in our model while being compatible with LFV and EDM bounds.

Concerning LFV processes, hereafter we focus only on $\mu \to e \gamma$ since it represents
the best probe of our scenario. The branching ratio of $\mu \to e \gamma$ is defined as
\begin{align}
{\rm BR}(\mu \to e \gamma) = \frac{48\pi^3 \alpha}{G^2_F} \left(\left|A^{21}_L\right|^2 + \left|A^{21}_R\right|^2 \right),
\end{align}
where the amplitudes $A^{21}_L$ and $A^{21}_R$, in the limit of $M_1 = M_2 = \mu = \tilde{m}_L = \tilde{m}_R = \tilde{m}$ and keeping only $\tan \beta$ enhanced 
terms,\footnote{We will eventually prefer low values for $\tan \beta$ ($\lesssim 5 $). For the numerical analysis later on one has therefore to take into account all 
contributions, which are collected in the Appendix. At this point we are rather interested in keeping the formulae simple and just give order-of-magnitude estimates.} 
read
\begin{eqnarray}
A^{21}_L & = & \frac{4\alpha_2 + 5\alpha_Y}{240\pi}~\frac{\tan\beta}{{\tilde m}^2}~(\delta_{LL}^e)_{21}
+ \frac{\alpha_Y}{48\pi}\left(\frac{{\tilde m}}{m_\mu}\right)
~\frac{1}{{\tilde m}^2}~(\delta_{LR}^e)^{*}_{12}\,,
\\
A^{21}_R & = & -\frac{\alpha_Y}{240\pi}~\frac{\tan\beta}{{\tilde m}^2}~(\delta_{RR}^e)_{21}
+ \frac{\alpha_Y}{48\pi}\left(\frac{{\tilde m}}{m_\mu}\right)
~\frac{1}{{\tilde m}^2}~(\delta_{LR}^e)_{21}\,.
\end{eqnarray}
Notice that in the above amplitudes we have kept only single MI effects since they are dominant
in our scenarios. The expressions for $\Delta a_\mu$ and $d_e$ are well approximated by
%
\begin{eqnarray}
\Delta a_\mu &=& \frac{5\alpha_2 + \alpha_Y}{48\pi}\frac{m^2_\mu}{{\tilde m}^2}\tan\beta\,,
\\
\frac{d_e}{e} &=& \frac{\alpha_Y}{120\pi}\frac{m_\tau}{{\tilde m}^2}\tan\beta~
{\rm Im} [(\delta_{LL}^e)_{13} (\delta_{RR}^e)_{31}]\,.
\end{eqnarray}
In order to highlight the relevant effects, we now provide some numerical
estimates for the above observables outlining also their possible correlations.
We find that
\begin{eqnarray}
\label{eq:muegamma}
{\rm BR}(\mu \to e \gamma) &\approx&
3 \times 10^{-14}\left(\frac{200~{\rm GeV}}{{\tilde m}}\right)^4
\tan^2\beta~\left(\frac{|(\delta_{LL}^e)_{21}|}{10^{-4}}\right)^2\,,
\\
\label{eq:delta_amu}
\Delta a_\mu &\approx& 3 \times 10^{-10} \left(\frac{200~{\rm GeV}}{{\tilde m}}\right)^2\tan\beta\,,
\\
\label{eq:de}
|d_e| &\approx& 2 \times 10^{-29} \left(\frac{200~{\rm GeV}}{{\tilde m}}\right)^2
\tan\beta~
\left|\frac{{\rm Im} [(\delta_{LL}^e)_{13} (\delta_{RR}^e)_{31}]}{10^{-6}}\right|e{\rm ~cm}\,.
\end{eqnarray}
Making the correlations among BR$(\mu \to e \gamma)$, $\Delta a_\mu$ and $d_e$ 
more explicit, it turns out that
\begin{eqnarray}
{\rm BR}(\mu \to e \gamma) &\approx& 3 \times 10^{-13}\left(\frac{\Delta a_\mu}{10^{-9}}\right)^2
\left(\frac{|(\delta_{LL}^e)_{21}|}{10^{-4}}\right)^2\,,
\\
|d_e| &\approx& 7 \times 10^{-29} \left(\frac{\Delta a_\mu}{10^{-9}}\right)
\left|\frac{{\rm Im} [(\delta_{LL}^e)_{13} (\delta_{RR}^e)_{31}]}{10^{-6}}\right|e{\rm ~cm}\,.
\end{eqnarray}

Eqs.~(\ref{eq:muegamma}-\ref{eq:de}) deserve few comments:
\begin{itemize}
\item In both flavor models we have considered, the dominant contribution to BR$(\mu\to e\gamma)$ stems 
from $A^{21}_L$, in particular from the $\tan\beta$-enhanced term proportional to $(\delta_{LL}^e)_{21}$, 
due to smaller flavor hierarchies in the left-handed lepton sector.
\item The dominant $\mu\to e\gamma$ amplitude grows with $\tan\beta$
as $A^{21}_L \sim (\delta_{LL}^e)_{21} \tan\beta \sim \tan^3\beta$, since
$(\delta_{LL}^e)_{21} \sim y^2_\tau \approx 10^{-4}\tan^2\beta$, which implies 
that $A^{21}_L$ is very efficiently suppressed for relatively low $\tan\beta$.
As a result, the very stringent experimental bound on BR$(\mu\to e\gamma)$
might be fulfilled even for a light spectrum ${\tilde m}\sim 200~$GeV
provided $\tan\beta\sim 1$.
\item The electron EDM can be induced at the leading order only through the effective
MI of Eq.~(\ref{triple_MI}) and it turns out that $d_e \sim \tan^5\beta$. Therefore $d_e$
is well under control for low $\tan\beta$ values, analogously to BR$(\mu\to e\gamma)$.
\item The $a_\mu$ anomaly can be accounted for while satisfying the stringent bounds from 
BR$(\mu \to e \gamma)$ and $d_e$,  only provided that the relevant flavor mixing angles 
are suppressed at the level of $(\delta_{LL}^e)_{21} \lesssim 10^{-4}$ and
$(\delta_{LL}^e)_{13} (\delta_{RR}^e)_{31} \lesssim 10^{-6}$.
\end{itemize}
In order to quantify the above considerations, we specialize now to the $U(1)$ flavor models
that have been introduced in the previous section: the {\it anarchical} and the {\it hierarchical} 
models. 
The predictions of other scenarios discussed in Refs.\cite{Altarelli:2012ia,Bergstrom:2014owa}, 
fall in between the ones we discuss here. In these two models, the relevant MIs entering 
the predictions of BR$(\mu \to e \gamma)$ and $d_e$ are estimated as:
\begin{itemize}
\item Anarchy
\begin{align}
(\delta_{LL}^e)_{21} & \approx  \kappa \frac{6}{N}~y^2_\tau 
\approx \kappa \frac{6 \times 10^{-4}}{N}\tan^2\beta,
\nonumber\\
(\delta_{LL}^e)_{13} (\delta_{RR}^e)_{31} & \approx 
\kappa' \frac{200}{N^2}~y^4_\tau \epsilon_A^{3} \approx 
\kappa' \frac{2 \times 10^{-8}}{N^2} \left(\frac{\epsilon_A}{0.2}\right)^3\tan^4\beta\,.
\label{eq:MI_anarchy}
\end{align}
\item Hierarchy
\begin{align}
(\delta_{LL}^e)_{21} & \approx \kappa \frac{6}{N}~y^2_\tau \epsilon^{3}_H 
\approx \kappa \frac{2 \times 10^{-5}}{N}\tan^2\beta \left( \frac{\eps_H}{0.3}\right)^3,
\nonumber\\
(\delta_{LL}^e)_{13} (\delta_{RR}^e)_{31} & \approx \kappa' \frac{200}{N^2}~y^4_\tau \epsilon_H^{7} 
\approx \kappa' \frac{4 \times 10^{-10}}{N^2} \left(\frac{\epsilon_H}{0.3}\right)^7\tan^4\beta\,.
\label{eq:MI_hierarchy}
\end{align}
\end{itemize}
where we have used Eqs.~(\ref{dLL_dRR}), (\ref{lErel}) assuming an intermediate mediation scale $M \sim 10^{10} \, \GeV$.
Moreover, we have explicitly included the dependence on the unknown ${\cal O}(1)$ coefficients 
parameterized through $\kappa$ and $\kappa'$ that are defined as 
\begin{align}
\label{kappadef}
\kappa & \equiv \frac{\kappa_{23} \kappa_{13}^*}{a_{33}^2}\,, &
\kappa' & \equiv \frac{\kappa_{13} \kappa_{31} \kappa_{33}^{*2} }{a_{33}^4}. 
\end{align}
A prominent feature emerging from Eqs.~(\ref{eq:MI_anarchy})-(\ref{eq:MI_hierarchy})
is the sensitivity of the MIs to the number of messenger $N$, since the 
diagonal sfermion masses are dominated by the MGM contribution proportional to $N$. 
We finally get for the $\mu \to e \gamma$ branching ratio  
\begin{align}
{\rm BR}(\mu \to e \gamma) &\approx
\tan^6\beta \left( \frac{\kappa}{N} \right)^2 \left(\frac{200~{\rm GeV}}{{\tilde m}}\right)^4
 \times  \begin{cases} 1 \times 10^{-12} & {\rm Anarchy} \\ 9 \times 10^{-16}  & {\rm Hierarchy} \end{cases} \\
 & \approx   \tan^4\beta \left( \frac{\kappa}{N} \right)^2 \left(\frac{\Delta a_\mu}{10^{-9}}\right)^2 \times \begin{cases} 1 \times 10^{-11} & {\rm Anarchy} \\ 9 \times 10^{-15}  & {\rm Hierarchy} \end{cases}
\end{align}
and the eEDM
\begin{align}
|d_e| &\approx
\tan^5\beta \left( \frac{\kappa'}{N^2} \right)  \left(\frac{200~{\rm GeV}}{{\tilde m}}\right)^2 e{\rm ~cm}\, \times  
\begin{cases} 
4 \times 10^{-31} & {\rm Anarchy} \\ 
1 \times 10^{-32}  & {\rm Hierarchy} 
\end{cases} \\
& \approx 
\tan^4\beta \left( \frac{\kappa'}{N^2} \right)  \left(\frac{\Delta a_\mu}{10^{-9}}\right) e{\rm ~cm} \times 
\begin{cases} 
1 \times 10^{-30} & {\rm Anarchy} \\ 
3 \times 10^{-32}  & {\rm Hierarchy} \,.
\end{cases}
\end{align}
Reformulating the constraint from $\mu \to e \gamma$ as a bound on the SUSY scale gives approximately
\begin{align}
\frac{\tilde{m}}{200 \, \GeV} \gtrsim \tan^{3/2} \beta \sqrt{\frac{\kappa}{N} } \left( \frac{{\rm BR}(\mu \to e \gamma) }{5.7 \times 10^{-13}} \right)^{1/4} \times \begin{cases} 1 & {\rm Anarchy} \\ 0.2  & {\rm Hierarchy}\,. \end{cases}
\end{align}
\begin{figure}[t!]
\centering
\includegraphics[scale=0.42]{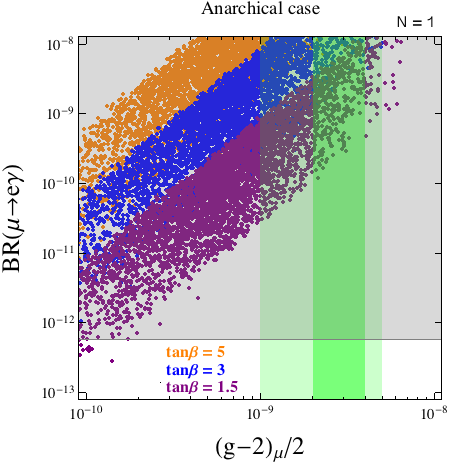}
 \hspace*{10pt}
\includegraphics[scale=0.42]{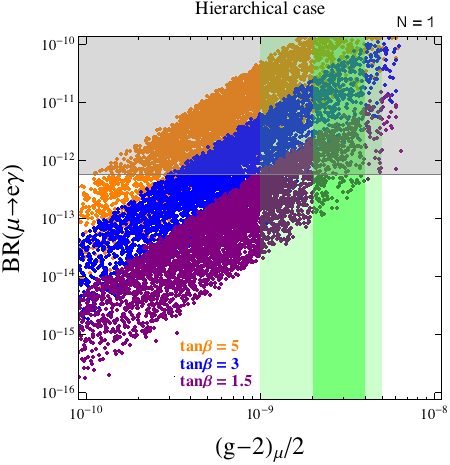}
\\ \vspace*{10pt}
\includegraphics[scale=0.42]{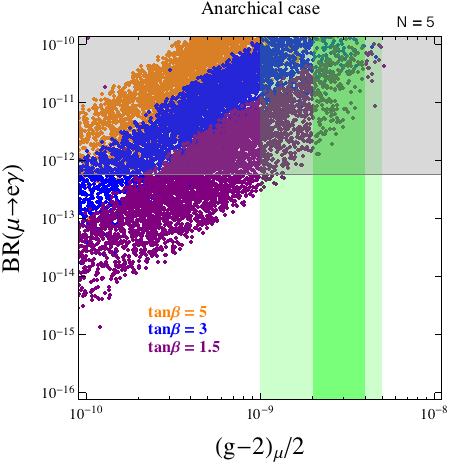}
\hspace*{10pt}
\includegraphics[scale=0.42]{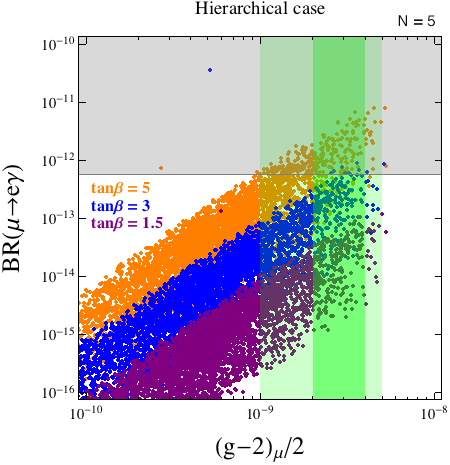}
\caption{
Predictions for BR$(\mu\to e\gamma)$ vs.~$\Delta a_\mu$ for different values of
$\tan\beta$: purple, blue and orange dots correspond to $\tan\beta = 1.5, 3, 5$,
respectively. The plots on the left (right) refer to the anarchical (hierarchical) case.
For the upper (lower) plots the number of messenger is set to $N = 1~(5)$.
}
\label{fig:amu_meg}
\end{figure}

Having outlined the expected behaviors and main features of flavor observables
within our FGM setup supplemented by $U(1)$ flavor models, we are ready now to
perform a complete numerical analysis.
\begin{figure}[t!]
\centering
\includegraphics[scale=0.42]{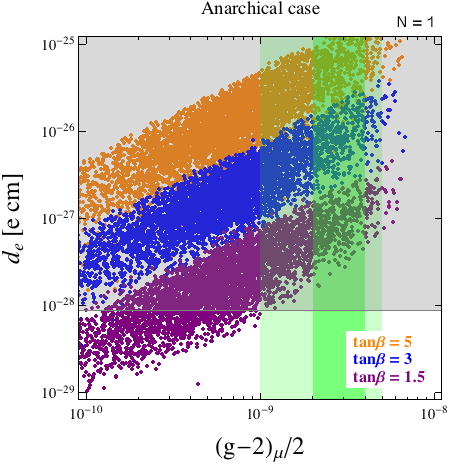}
\hspace*{10pt}
\includegraphics[scale=0.42]{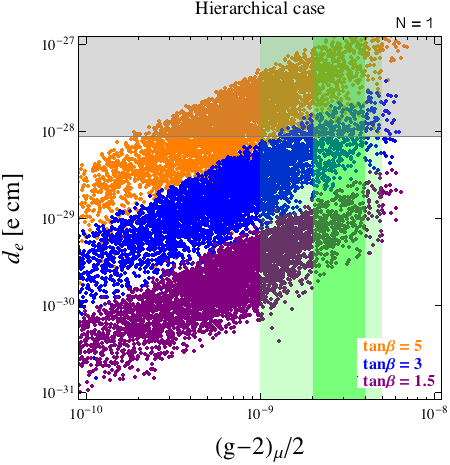}
\\ \vspace*{10pt}
\includegraphics[scale=0.42]{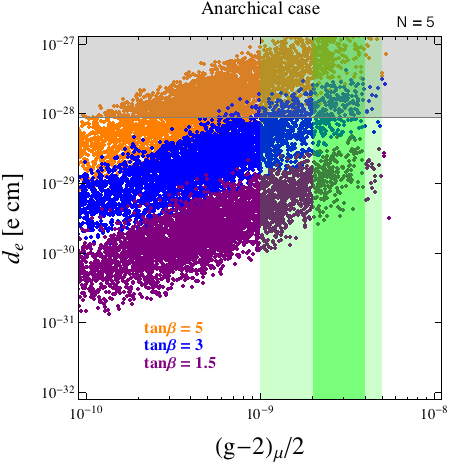}
\hspace*{10pt}
\includegraphics[scale=0.42]{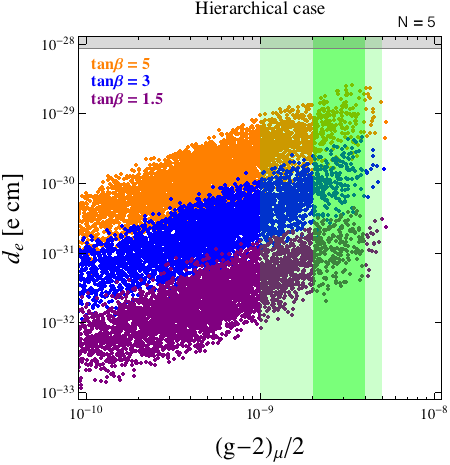}
\caption{
The same scenarios as in Fig.~\ref{fig:amu_meg} for $d_e$ vs.~$\Delta a_\mu$.
}
\label{fig:amu_de}
\end{figure}
\begin{figure}[t!]
\centering
\includegraphics[scale=0.42]{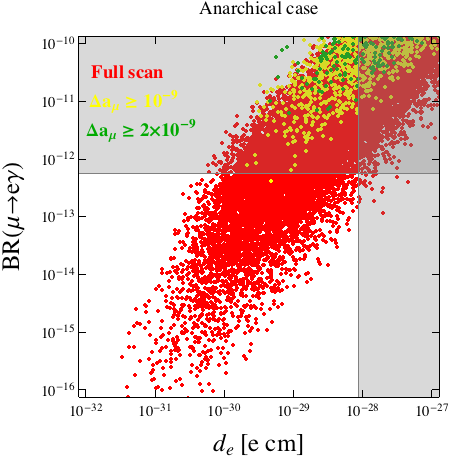}
\hspace{10pt}
\includegraphics[scale=0.42]{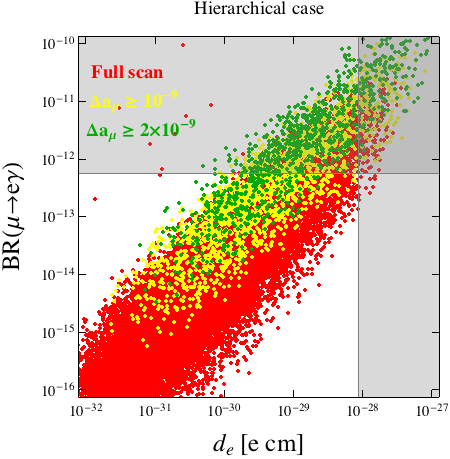}
\caption{Predictions for BR$(\mu\to e\gamma)$ vs.~$d_e$ for the anarchical (left) and hierarchical (right) 
cases.  Yellow (green) points correspond to $\Delta a_\mu \ge 10^{-9}$ ($2\times 10^{-9}$). 
}
\label{fig:summary}
\end{figure}
In Fig.~\ref{fig:amu_meg}, we show the predictions for BR$(\mu\to e\gamma)$
vs. $\Delta a_\mu$ for different values of $\tan\beta$: purple, blue and orange
dots correspond to $\tan\beta = 1.5, 3, 5$, respectively. The plots on the left
(right) refer to the anarchical (hierarchical) case. For the upper (lower) plots
the number of messenger is set to $N = 1~(5)$.
In Fig.~\ref{fig:amu_de}, we show the analogous plots for $d_e$ vs. $\Delta a_\mu$.
In the scan we have varied the unknown $\mathcal{O}(1)$ coefficients $\kappa, \kappa'$ for
the MIs in the range $(0.3, 1.5)$. 
The other parameters were varied in the following ranges:
\begin{align}
10^6 ~{\rm GeV}\le M \le 10^{15} ~{\rm GeV},~~ 100~{\rm GeV} \le \tilde{m}_{E}(M) \le 1~{\rm TeV},  ~~
100~{\rm GeV} \le \mu \le \mu_{\rm max},
\end{align}
where $\mu_{\rm max} \equiv \tilde{m}_{\tau_L} \tilde{m}_{\tau_R}/(m_\tau \tan\beta)$ is the maximal 
value giving a non-tachyonic stau.
In the plots, the gray shaded regions are excluded by the current bounds from $\mu\to e\gamma$ or $d_e$ 
reported in Table~\ref{tab:LFV}, while the green (dark green) area approximately corresponds to values of 
$\Delta a_\mu$ lowering the discrepancy below the 2$\sigma$ (1$\sigma$) level.

A direct comparison of the bounds and the discovery potential of  $\mu\to e\gamma$ and $d_e$
is shown in Fig.~\ref{fig:summary}, where we plot the result of a random variation of the full set 
of parameters for the anarchical (left) and hierarchical (right) cases:
\begin{gather}
 10^6 ~{\rm GeV}\le M \le 10^{15} ~{\rm GeV},~~ 100~{\rm GeV} \le \tilde{m}_{E}(M) \le 2~{\rm TeV},  ~~
100~{\rm GeV} \le \mu \le \mu_{\rm max}, \nonumber \\
 1 \le N \le 5,\quad 1.5 \le \tan\beta \le 5,\quad 0.3\le \kappa, \kappa^\prime \le 1.5. 
\end{gather}
In addition, the yellow (green) points correspond to $\Delta a_\mu \ge 10^{-9}$ ($2\times 10^{-9}$).

The main results emerging from our numerical analysis can be summarizes as follows:
\begin{itemize}
\item In the anarchical scenario, it is very difficult if not impossible to explain
the $\Delta a_\mu$ anomaly because of the strong bounds from both BR$(\mu\to e\gamma)$
and $d_e$ (see left panels of Figs.~\ref{fig:amu_meg}-\ref{fig:summary}). 
The latter observables have a comparable sensitivity to the scenario in
question and might reach experimentally visible values even for SUSY masses far beyond
the LHC reach in the multi-TeV regime. As already discussed, BR$(\mu\to e\gamma)$ and
$d_e$ grow fast with $\tan\beta$ (like $\tan^6\beta$ and $\tan^5\beta$, respectively)
and are both suppressed by increasing $N$. As a result, the scenario with low $\tan\beta$
and $N=5$ (we remind that for low mediation scales perturbativity requires $N \lesssim 5$) 
is the most favorable scenario, as clearly shown in Figs.~\ref{fig:amu_meg} and \ref{fig:amu_de}.
\item The hierarchical scenario easily offers the possibility to explain the
$\Delta a_\mu$ anomaly while satisfying the limits on BR$(\mu\to e\gamma)$ and $d_e$
(right panels of Figs.~\ref{fig:amu_meg}-\ref{fig:summary}).
This happens thanks to the stronger suppression of the flavor mixing angles compared
to the anarchical case. On the other hand, all the others considerations made above for
the anarchical case apply here as well.
\item 
$\mu\to e\gamma$ and $d_e$ have comparable sensitivities, but $\mu\to e\gamma$ is currently more constraining,
as we can see from Fig.~\ref{fig:summary}. 
Interestingly,  an improvement of the sensitivity by one or two orders of magnitude would make the electron EDM 
the most powerful probe of FGM scenarios especially in case of heavy superpartners, corresponding to the red
points in Fig.~\ref{fig:summary}.
This is a consequence of the slower decoupling of $d_e$ with respect to the 
NP scale: $d_e \sim {\tilde m}^{-2}$, while BR$(\mu\to e\gamma)\sim {\tilde m}^{-4}$.
\item Given the expected future sensitivities to the $\mu\to e $ transitions reported in
Table~\ref{tab:LFV} and the following approximate relations among different decay modes:
\begin{eqnarray}
{{\rm BR}(\mu\to eee)}
&\simeq&
\frac{\alpha}{3\pi}
\bigg(\log\frac{m^2_{\mu}}{m^2_{e}}-3\bigg)
{\rm BR}(\mu\to e\gamma)~,
\nonumber \\
{\rm CR}(\mu\to e~\mbox{in N})
&\simeq&
\alpha\times {\rm BR}(\mu\to e\gamma)~,
\label{eq:dipole}
\end{eqnarray}
we see that there are good prospects for a full test of the parameter space favored by 
$\Delta a_\mu$ at future experiments.
\end{itemize}
Let us now also show how the $\mu\to e\gamma$ and $d_e$ constraints appear in terms of the gaugino and slepton masses. 
For illustration purposes, we adopt a more general low-energy spectrum than the one predicted by MGM, which allows us 
to parameterize in a model-independent way possible distortions of the spectrum due to the full set of matter-messenger 
couplings studied in \cite{Evans:2013kxa}, including the other couplings in Eq.~(\ref{eq:W}), as well as more generic 
SUSY breaking sectors, in the spirit of General Gauge Mediation \cite{Meade:2008wd}. 
In practice, we still use Eqs.~(\ref{eq:m2L}-\ref{eq:Ae})
to set the off-diagonal entries but we treat slepton and gaugino masses as free parameters at low energy. 
\begin{figure}[t!]
\centering
\includegraphics[scale=0.42]{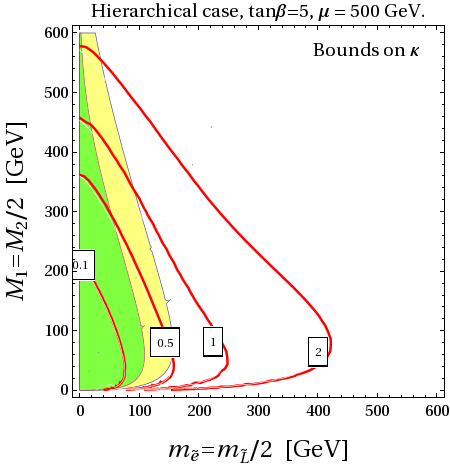}
\hspace{10pt}
\includegraphics[scale=0.42]{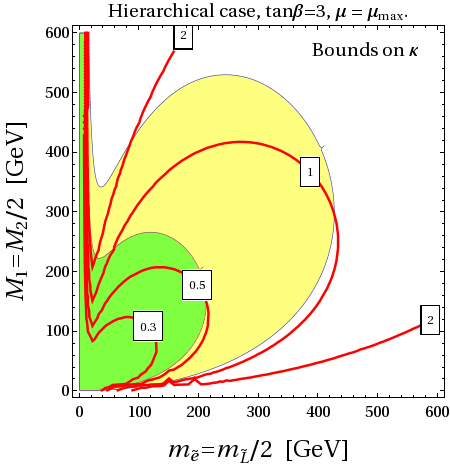}
\caption{
Bounds on the $\mathcal{O}(1)$ coefficient $\kappa$, see Eq.~(\ref{kappadef}), from $\mu\to e\gamma$ in the
hierarchical scenario. For definiteness, we have assumed $M_2 = 2 M_1$, $\tilde{m}_L = 2 \tilde{m}_E$ and
different choices of $\mu$ and $\tan\beta$. The yellow (green) areas give $\Delta a_\mu \ge 10^{-9}$ ($2\times10^{-9}$). 
}
\label{fig:contours1}
\end{figure}
\begin{figure}[t!]
\centering
\includegraphics[scale=0.42]{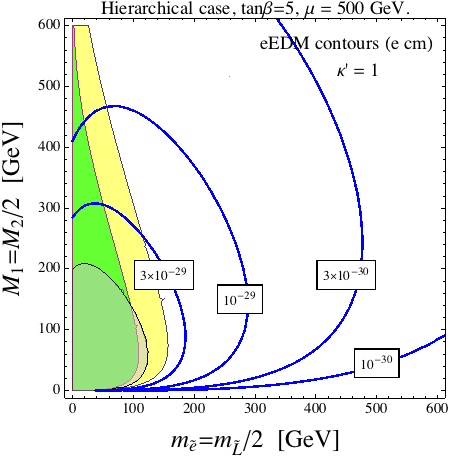}
\hspace{10pt}
\includegraphics[scale=0.42]{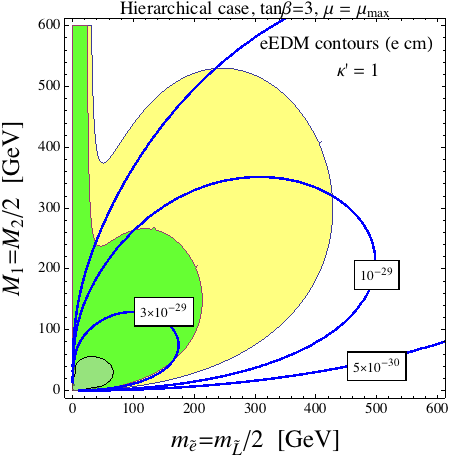}
\\ \vspace*{10pt}
\includegraphics[scale=0.42]{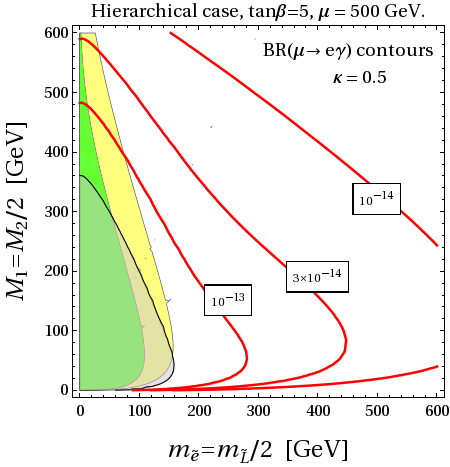}
\hspace{10pt}
\includegraphics[scale=0.42]{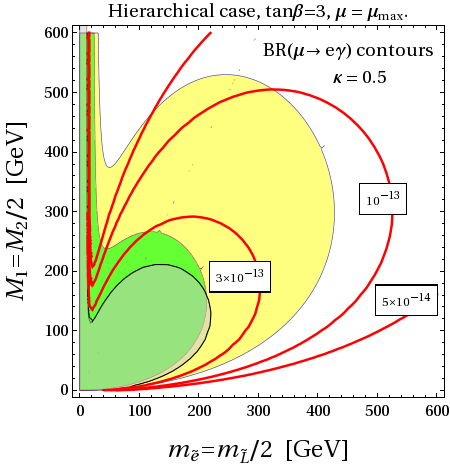}
\caption{
Contours of BR($\mu\to e\gamma$) and $d_e$ for the same choice of the parameters as in Fig.~\ref{fig:contours1} 
and $\kappa=0.5$, $\kappa'=1$. The gray shaded regions are presently excluded by $\mu\to e\gamma$ or $d_e$.
}
\label{fig:contours}
\end{figure}

In Fig.~\ref{fig:contours1}, we show the current bounds on the $\mathcal{O}(1)$ coefficient $\kappa$, as defined in Eq.~(\ref{kappadef}), 
from $\mu\to e\gamma$ for different choices of $\mu$ and $\tan\beta$ in the hierarchical case. For definiteness, we fixed the relation
among gauginos and slepton masses as follows: $M_2 = 2\times M_1$, $\tilde{m}_L = 2\times \tilde{m}_E$. The yellow (green) areas give 
$\Delta a_\mu \ge 10^{-9}$ ($2\times10^{-9}$). As we can see, it is not necessary that the unknown coefficients conspire to provide an unnaturally small suppression, in order to take $\mu\to e\gamma$  under control in the region favored by $(g-2)_\mu$.

In Fig.~\ref{fig:contours}, we show contours of BR($\mu\to e\gamma$) and $d_e$ for the same choice of the parameters as above and
the specific values $\kappa=0.5$, $\kappa'=1$, in order to illustrate the present bounds and the possible impact of the future experiments 
in terms of the masses of the SUSY particles in the game. In particular, the gray shaded regions are presently excluded by 
$\mu\to e\gamma$ or $d_e$. Again, we see that a large contribution to $(g-2)_\mu$ is perfectly compatible with the present bounds,
but there are good prospects for a full test of the relevant parameter space in the future.

Besides the LEP constraints (corresponding to $\tilde{m}_E,~\tilde{m}_L, M_2\gtrsim 100~{\rm GeV}$),
the mass plane shown in the above plots is now challenged by searches for the electroweak production of SUSY particles
performed by the LHC experiments, based on events with two or more leptons plus missing transverse momentum.
The exact bounds are model dependent and their precise derivation is beyond the scope of the present study. Nevertheless,
we briefly summarize here their possible impact. 

Since we are considering scenarios with gauge-mediated SUSY breaking, the LSP is always a practically massless gravitino.
The limits set by LHC searches then strongly depend on the nature and the life-time of the next to LSP (NLSP).
In case the sleptons are lighter than the Bino, as it can occur even in MGM for large values of $N$,
each decay chain would end with the degenerate sleptons NLSP decaying into leptons and gravitino. If this
decay occurs promptly (which requires low mediation scales, $M\lesssim 10^6$ GeV), 
then the bounds from direct (Drell-Yan) slepton production translate to a limit on the mass of the right-handed (left-handed) sleptons at about 250 (300) GeV \cite{Aad:2014vma}. 
In case sleptons are long-lived compared to the detector scale, searches for charged tracks set a bound on 
degenerate slepton NLSP mass at about 400 GeV \cite{Chatrchyan:2013oca}. 
Interestingly, no searches performed so far constrain the intermediate case, featuring a disappearing track with a displaced vertex inside the detector, occurring for a wide range of the messenger scale, $10^6~{\rm GeV} \lesssim M \lesssim10^9~{\rm GeV}$ \cite{CMPR}.
The above limits can be substantially relaxed if the Bino is lighter than the sleptons and escapes the detector, thus resembling 
searches within gravity mediation. In particular for a neutralino NLSP heavier than about 150 GeV, there is no constraint 
from direct slepton production  \cite{Aad:2014vma}.

The most stringent constraint would occur in the case of the hierarchy $M_1 < \tilde{m}_L < M_2$ from Wino-like
chargino/neutralino production followed by decays into on-shell sleptons/sneutrinos, with bounds up to 700 GeV on the Wino
mass from multi-lepton plus missing energy searches \cite{Khachatryan:2014qwa}. However, such searches loose sensitivity if the mass splitting
of the sleptons with either the Bino or the Wino gets small. 

Comparing the limits reported above, with our plots in Fig.~\ref{fig:contours1} and \ref{fig:contours}, 
we see that there is still room for a large SUSY contribution to $a_\mu$, at least at the $10^{-9}$ level (yellow regions),
especially if $\mu$ approaches the maximal value  $\mu_{\rm max} \equiv  \tilde{m}_{\tau_L} \tilde{m}_{\tau_R}/(m_\tau \tan\beta)$.
This conclusion is supported by the results of Refs.~\cite{Endo:2013bba,Endo:2013lva} where a systematic study 
of the LHC constraints on the parameter space favored by $(g-2)_\mu$ has been presented.

\section{Comparison with other models}

In this section, we compare the peculiar flavor structure of FGM to other models that predict
the parametric flavor suppression of soft terms. In particular, we consider $U(1)$ flavor 
models within SUGRA scenarios and models with SUSY Partial Compositeness (PC).

In those models the SUSY mediation scale $\Lambda_S$ is assumed to be above the scale
of flavor messengers $\Lambda_F$, so that the flavor structure of soft terms at the scale 
$\Lambda_F$ is controlled entirely by the flavor dynamics at this scale, irrespectively
of their structure at the scale $\Lambda_S$. In FGM the situation is reversed as the SUSY
messenger scale $\Lambda_S = M$ is below $\Lambda_F$. We stress that this setup 
is therefore complementary to the other scenarios, allowing also for very low SUSY mediation scales. 
All the unspecified dynamics of the flavor sector is imprinted in the matter-messenger couplings, 
just like Yukawas, and the full SUSY spectrum is totally calculable in terms of these couplings.

\subsection{$U(1)$ and Gravity Mediation}
In Gravity Mediation the natural expectation for soft terms at the flavor scale is given
by the most general terms invariant under the symmetry using the flavon as a spurion.
This gives for slepton mass insertions
\eq{(\delta_{LR}^e)_{ij}  \sim \frac{A v_d}{\tilde{m}_L \tilde{m}_E }  \eps^{L_i + E_j}\,,}
\begin{align}
 (\delta_{LL}^e)_{ij} & \sim\epsilon^{|L_i - L_j |}\,, &
   (\delta_{RR}^e)_{ij} & \sim \epsilon^{|E_i - E_j |}\,.
\end{align}
Focusing on the {\it anarchical} and {\it hierarchical} models of Section 3.1, the relevant MIs for
BR$(\mu\to e\gamma)$ and $d_e$ are again $(\delta_{LL}^e)_{21}$ and $(\delta_{LL}^e)_{13}(\delta_{RR}^e)_{31}$, 
respectively. In the anarchic case we find:
\begin{itemize}
\item Anarchy
\begin{eqnarray}
(\delta_{LL}^e)_{21} \sim 1 \,,\qquad
(\delta_{LL}^e)_{13}(\delta_{RR}^e)_{31} \sim \epsilon^3_A \,,
\label{eq:sugra_anarchy}
\end{eqnarray}
leading to the following predictions:
\begin{eqnarray}
\label{eq:muegamma_sugra_ana}
{\rm BR}(\mu \to e \gamma) &\sim&
5 \times 10^{-13}\left(\frac{10~{\rm TeV}}{{\tilde m}}\right)^4
\tan^2\beta\,,
\\
\label{eq:de_sugra_ana}
|d_e| &\sim& 7 \times 10^{-29} \left(\frac{10~{\rm TeV}}{{\tilde m}}\right)^2
\tan\beta \, e {\rm \, cm}\,.
\end{eqnarray}
\item Hierarchy
\begin{eqnarray}
(\delta_{LL}^e)_{21} \sim \epsilon_H \sim 0.3 \,,\qquad
(\delta_{LL}^e)_{12}(\delta_{RR}^e)_{21} \sim \epsilon^3_H\,,
\label{eq:MI_hierarchyGrav}
\end{eqnarray}
where we took $(\delta_{LL}^e)_{12}(\delta_{RR}^e)_{21}$ instead of
$(\delta_{LL}^e)_{13}(\delta_{RR}^e)_{31}$ since the contribution of the latter to 
the eEDM is smaller by a factor of $(y_{\tau}/y_{\mu})\times \epsilon^{4}_{H} \sim 0.1$
compared to that induced by $(\delta_{LL}^e)_{12}(\delta_{RR}^e)_{21}$. 
We therefore have the following predictions:
\begin{eqnarray}
\label{eq:muegamma_sugra_hie}
{\rm BR}(\mu \to e \gamma) &\sim&
7 \times 10^{-13}\left(\frac{5~{\rm TeV}}{{\tilde m}}\right)^4
\tan^2\beta~\left(\frac{\epsilon_H}{0.3}\right)^2\,,
\\
\label{eq:de_sugra_hie}
|d_e| &\sim& 6 \times 10^{-29} \left(\frac{5~{\rm TeV}}{{\tilde m}}\right)^2
\tan\beta \, e {\rm \, cm}\,.
\end{eqnarray}
\end{itemize}
As a result, single $U(1)$ flavor models with Gravity Mediation need sleptons well above the TeV scale, 
${\tilde m} \gtrsim 10~{\rm TeV}\times \sqrt{\tan\beta} ~(5~{\rm TeV}\times \sqrt{\tan\beta})$ in the 
anarchical (hierarchical) scenario. Note that the bounds on the SUSY spectrum in the quark 
sector, in particular from $\eps_K$, are much stronger~\cite{Calibbi:2013mka}.
\begin{table}
\centering
\begin{tabular}{c|c|c||c}
& SUGRA $U(1)$ & PC & FGM $U(1)$\\
\hline
\hline
& & & \\
$\frac{\tilde{m}}{m_e} {\rm Im} (\delta_{LR})_{11}$ & $1$ & $1$ & $ y_\tau^4 t_\beta $ \\
& & & \\
$\frac{\tilde{m}}{m_\mu} (\delta_{LR})_{12}$ & $\eps^{L_{1}-L_{2}}$ & $ \eps^{L_{1}-L_{2}}$ &   $y_\tau^2 \eps^{L_{1}+L_{2}-2 L_{3}}$ \\
& & & \\
$\frac{\tilde{m}}{m_\mu} (\delta_{LR})_{21}$ & $\eps^{E_1 - E_2}$ & $ \eps^{E_1 - E_2}$ &  $y_\tau^2 \eps^{ E_1 - E_2 + 2 L_1 - 2 L_3}$ \\
& & &  \\
$(\delta_{LL})_{12}$ & $\eps^{L_{1} - L_{2}}$ & 
$ \eps^{L_{1} + L_{2}}$ 
& $y_\tau^2 \epsilon^{L_{1} + L_{2} -2 L_{3}}$ \\
& & & \\
$(\delta_{RR})_{12}$ & $\eps^{E_1 - E_2}$ & 
$ \eps^{E_{1} + E_{2}}$ 
& $y_\tau^2 \epsilon^{E_1 + E_2 - 2E_3}$
\end{tabular}
\caption{Predictions for the mass insertions in various SUSY models with an underlying $U(1)$ flavor model where $L_i$ ($E_i$) stands for the charges of $SU(2)$ doublets (singlets). Note that for the sake of simplicity  we compare only single mass insertions, for large $\tan \beta $ triple mass insertions can possibly give the dominant contributions to LR transitions.}
\label{deltaijSU5}
\end{table}

\subsection{SUSY Partial Compositeness}

According to the paradigm of Partial Compositeness, the lepton Yukawa matrices have the form
\begin{equation}
\label{eq:yukawas}
(y_E)_{ij} \sim g_{\rho} \epsilon^\ell _i \epsilon^e_j,
\end{equation}
where $g_\rho$ is a strong coupling and $\epsilon^{\ell,e}_i \lesssim 1$ measures the amount
of compositeness for the leptons. Such a scheme closely resembles the case of a single $U(1)$ 
flavor model, with the correspondence (in the limit of $g_\rho = 1$)
\be
\epsilon^{\ell,e} _i \, \longleftrightarrow \, \epsilon^{L_i,E_i}\,.
\label{eq:U(1)_vs_PC}
\ee
As a result, the MIs are expected to take the following form~\cite{Lodone}
\be
(\delta_{LL}^e)_{ij} \sim \eps^\ell_i \eps^\ell_j \sim \eps^{L_i + L_j },
\qquad
(\delta_{RR}^e)_{ij} \sim \eps^e_i \eps^e_j \sim  \eps^{E_i + E_j }.
\ee
%
%
\be
(\delta_{LR}^e)_{ij} \sim \frac{v_d A g_\rho}{\tilde{m}_L \tilde{m}_E}
\eps^\ell_i \eps^e_j \sim \frac{m^e_i A}{\tilde{m}_L \tilde{m}_E}\eps^{E_j - E_i} 
\sim \frac{m^e_j A}{\tilde{m}_L \tilde{m}_E}\eps^{L_i - L_j} \,.
\ee
%
%
%
In PC, the leading contributions to BR$(\mu\to e\gamma)$ typically arise from
$(\delta_{LR}^e)_{12}$. In particular, in the anarchical scenario we find
\begin{eqnarray}
\label{eq:muegamma_sugra_ana}
{\rm BR}(\mu \to e \gamma) &\sim&
6 \times 10^{-13}\left(\frac{5~{\rm TeV}}{{\tilde m}}\right)^4\,,
\end{eqnarray}
while in the hierarchical case we have a mild additional suppression by a factor of $\epsilon^2_H \approx 0.1$.
Note however that in PC the left-handed ``charges" $L_i$ are determined from the PMNS matrix analogously to $U(1)$ 
models only in the case of light Dirac Neutrinos. If instead light neutrinos are Majorana, then the Weinberg operator 
can arise from a bilinear coupling to the composite sector (instead of linear couplings that resemble the $U(1)$ 
structure). In this case only the combination $L_i + E_j$ is determined by charged lepton Yukawa couplings, and 
the constraints from LFV can be significantly relaxed by choosing symmetric charges~\cite{Lodone} 
\begin{align}
\eps^{L_i} \sim \eps^{E_i} \sim \sqrt{\frac{y_i^e}{g_\rho}}.
\end{align}   
This implies
\begin{align}
\frac{\tilde{m}}{m_\mu} (\delta_{LR}^e)_{12} \sim \eps^{L_1 - L_2} \sim 
\sqrt{\frac{m_e}{m_\mu}}\,,
\end{align}
%
%
and thus
\begin{align}
{\rm BR}(\mu \to e \gamma) \sim 7 \times 10^{-13}\left(\frac{1.5~{\rm TeV}}{{\tilde m}}\right)^4\,.
\end{align}
On the other hand, the predictions for the electron EDM are completely independent of any 
charge assignments since in PC the diagonal elements of the A-terms are generally complex. 
As a result, we find
\begin{eqnarray}
|d_e| &\sim& 7 \times 10^{-29} \left(\frac{3~{\rm TeV}}{{\tilde m}}\right)^2 
{\rm Im} \left( \frac{M_1 A}{\tilde{m}^2}\right)~{\rm e~cm}\,,
\end{eqnarray}
and therefore the eEDM now provides the strongest constraint on the PC scenario. Notice that the electron EDM
has a similar sensitivity to NP effects in PC scenarios and $U(1)$ flavor models, independently of 
the particular charge assignments, pushing the SUSY scale to ${\tilde m} \gtrsim 3 \div 5 ~{\rm TeV}$. 
Needless to say, neither  PC scenarios nor SUGRA with an underlying $U(1)$ flavor model can explain 
the muon $g-2$ anomaly.

In Table~\ref{deltaijSU5}, we summarize the predictions for the MIs most relevant
for phenomenology in various models: SUGRA (first column), PC (second column) and 
FGM (last column).
On general ground, comparing the flavor structure of the soft sector of SUGRA and
PC/FGM scenarios, the most prominent feature is the higher suppression for off-diagonal
sfermion masses in the LL and RR sectors in the PC/FGM case. The LR sector has the
same parametric structure in PC and SUGRA, since in both scenarios the A-terms are
proportional to the SM Yukawas, while in FGM we have a much stronger suppression
arising from a partial alignment among SM Yukawas and A-terms.
Finally, PC and SUGRA share also the same SUSY CP problem as they allow complex diagonal
elements for the A-terms. In contrast, within FGM, the leading CPV phases arise only at
higher order in MI expansions and therefore are very suppressed.

\section{Conclusions}

Now that the Higgs boson has been discovered, naturalness becomes a pressing question waiting 
for the final answer of LHC14. If new dynamics is present around the TeV scale, as needed to 
explain the smallness of the electro-weak scale, one would expect too large contributions 
to flavor transitions mediated by the new physics states unless some protection mechanism is at work. 
Therefore, the possibility of finding new physics at the LHC is closely related to the existence 
of a suppression of flavor violating processes.

In this respect, Minimal Gauge Mediation (MGM) provides an ideal framework to
accomplish this job. Indeed, if the flavor scale is much higher than the SUSY messenger
scale then the flavor structure of the soft terms is entirely determined by the SM Yukawas, thus realizing the
paradigm of Minimal Flavor Violation (MFV)~\cite{D'Ambrosio:2002ex}. The drawback of this
scenario is that any imprint of the flavor sector in low-energy physics and thus the
possibility to test the flavor dynamics is completely lost.

On the other hand, minimal realizations of GMSB are now seriously challenged by the Higgs
boson discovery at the LHC, since they can account for $m_h\approx 126$ GeV only at the
price of a SUSY spectrum that is beyond the reach of the LHC.
This has motivated extensions of minimal GMSB models by introducing new direct couplings between the messengers and the MSSM matter fields in order to obtain a large Higgs mass for
light stops by generating non-vanishing A-terms at the messenger scale~\cite{Evans:2011bea, Draper:2011aa, Evans:2012hg, Kang:2012ra, Craig:2012xp, Albaid:2012qk, Abdullah:2012tq, Byakti:2013ti, Evans:2013kxa, Jelinski:2013kta, Jelinski:2014uba}.

Among these scenarios, ``Flavored Gauge Mediation" (FGM)~\cite{Shadmi:2011hs} assumes that
these new couplings have a flavor structure which is controlled by the same underlying
flavor symmetry that explains the smallness of Yukawa couplings. As a result, FGM allows
to generate soft masses which still carry information about the high scale flavor sector.
Interestingly, due to the loop origin of soft terms, sfermion masses exhibit a flavor
pattern that is much stronger suppressed than in Gravity Mediation.
This strong suppression, arising even in the context of single $U(1)$ flavor models, is
reminiscent of what happens in the case of wave function renormalization~\cite{Isidori,Pokorski}
or Partial Compositeness~\cite{Nomura,Lodone}. In addition there is a strong suppression of LR flavor transitions that is particularly effective for accompanying flavor-blind phases,  thus rendering the strong bounds from EDMs under control. Therefore FGM does not only modify the SUSY spectrum of MGM in a way interesting for collider phenomenology, but it also allows to obtain a rich flavor phenomenology beyond MFV. In particular it offers a viable SUSY implementation of simple U(1) flavor symmetry models that in the context of gravity mediation have huge difficulties in passing the bounds from precision observables, and in the context of MGM are not testable at all. 

While in Ref.~\cite{Calibbi:2013mka} we concentrated on the quark sector, in this work we have focused on the lepton sector analyzing the implications of FGM with underlying $U(1)$ flavor models. In particular, we have studied the predictions of two models ({the \it anarchical} and {\it hierarchical} scenarios of Ref.~\cite{Altarelli:2012ia}) that are representative for a whole class of $U(1)$ models that accommodate lepton masses and mixing angles. We have analyzed $\mu\to e \gamma$ (which
turned out to be the most constraining LFV channel), the electron EDM $d_e$ and the muon anomalous magnetic moment $\Delta a_\mu$.
Since the experimental bounds on both $\mu \to e \gamma$ and $d_e$ underwent recently an impressive improvement, an important question of this work was to establish whether and to which extent single $U(1)$ flavor models were viable in the context of FGM models. A related relevant question was to establish whether the current muon $g-2$ anomaly
could be resolved while being compatible with the LFV and EDM bounds.

In the following, we summarize our main findings:
%
%
\begin{itemize}
\item The non-holomorphic soft masses (LL and RR mass insertions) are suppressed by powers
of the spurion that are the {\it sum} of U(1) charges, in contrast to the corresponding
gravity-mediated case where charge differences enter. The origin of this suppression is due
to the fact that the $U(1)$ controls soft terms only indirectly via the messenger sector,
which in turn generates soft terms only at loop level, thus leading to a double suppression
by small couplings. This strong suppression is similar to the cases of wave function renormalization or SUSY Partial Compositeness.
\item The A-terms are much more suppressed than in PC and gravity-mediated scenarios
where they are proportional to the leading order term $\epsilon^{L_i + E_j}$ allowed
by the $U(1)$ symmetry. Moreover, the A-terms are partially aligned to the Yukawa
couplings and their diagonal components are real therefore not inducing contributions
for the EDM. The first non-vanishing CP violating phase in the diagonal A-terms can
arise only through higher order expansions in the MIs, which leads to am additional suppression by powers of $y_\tau$ that becomes particularly effective for low $\tan \beta$.   
%
%
%
 \item LFV processes and the electron EDM can be kept under control even for a light
spectrum well below the TeV scale, provided that $\tan \beta$ is small (the smaller the better). This is true both in the {\it anarchical} and 
especially in the {\it hierarchical} scenarios. In contrast, PC and gravity-mediated
scenarios require a very heavy spectrum above the TeV scale in order to fulfill the experimental
bounds on BR$(\mu\to e\gamma)$ and $d_e$. Very low values for $\tan \beta$ also perfectly fit a complete realization of this setup within the NMSSM, which is the natural choice to generate the $\mu$-term in FGM scenarios, solving also the $\mu-B_\mu$ problem of MGM~\cite{Craig:2012xp, new}.
\item In spite of the tremendous experimental bounds on the electron EDM and LFV processes,
we have found that is still possible to account for the muon $g-2$ anomaly within FGM models
in the {\it hierarchical} but not in the {\it anarchical} scenarios. The same conclusion is
not true in PC and gravity-mediated models where both $\mu\to e\gamma$ and $d_e$ prevent
any sizable effect in $\Delta a_\mu$.
\item Although BR$(\mu\to e\gamma)$ and $d_e$ have comparable sensitivities to FGM scenarios,
$\mu\to e\gamma$ is currently more constraining. However, considering the slower decoupling of
NP effects in $d_e \sim {\tilde m}^{-2}$ with respect to BR$(\mu\to e\gamma)\sim {\tilde m}^{-4}$,
the electron EDM might become the most powerful probe of the scenarios in question
with improved experimental data especially in case of heavy superpartners.
\end{itemize}
In conclusion, we have analyzed in detail the anatomy and phenomenology in the lepton
sector of FGM models with underlying $U(1)$ models. Remarkably, it turned out that these models can pass the impressive bounds
on LFV processes and leptonic EDMs even for light superpartners, potentially observable at
LHC14, leaving open the possibility of accommodating the longstanding muon $(g-2)$ anomaly and testing $U(1)$ flavor models in upcoming experiments.

\section*{Acknowledgments}
The work of P.P. was supported by the ERC Advanced Grant No.267985 {\em Electroweak Symmetry 
Breaking, Flavour and Dark Matter: One Solution for Three Mysteries (DaMeSyFla)}.
This work made in the ILP LABEX (under reference ANR-10-LABX-63) was partially supported by French state funds
managed by the ANR within the Investissements d'Avenir programme under reference ANR-11-IDEX-0004-02.


\appendix

\section{Formulae for the leptonic dipoles}

In this appendix we collect the formulae for LFV branching ratios BR$(l_i \to l_j \gamma)$, lepton anomalous magnetic moments $\Delta a_i$, and lepton electric dipole moments $d_i$. The results have been obtained from the exact results of Refs.~\cite{Hisano, Moroi} using a generalized mass insertion approximation (see e.g.~Ref.~\cite{Paride}) without assuming large $\tan \beta$.   

\subsection{Lepton Flavor Violation: $l_i \to l_j \gamma$}
The branching ratio ${\rm BR}(l_i \to l_j \gamma)$ is given by 
\begin{align}
\frac{{\rm BR}(l_i \to l_j \gamma)}{{\rm BR}(l_i \to l_j \nu_i \overline{\nu}_j)} = 
\frac{48 \pi^3 \alpha}{G_F^2} \left( |A^{ij}_{L}|^2 + |A^{ij}_{R}|^2 \right).
\end{align}
The amplitude $A^{ij}_L$ receives a neutralino and chargino contribution $A^{ij}_L = A^{ij(n)}_L + A^{ij(c)}_L$, 
whereas $A^{ij}_R$ receives only a neutralino contribution $A^{ij}_R = A^{ij(n)}_R$. These contributions are given by:
\begin{align}
A^{ij(n)}_L & = 
\frac{\alpha_2}{4 \pi} 
\frac{(m_{LL}^2)_{ij}}{m_L^4} \left[ f_{1n} (x_{2L}) + \frac{(|M_2|^2 + \mu M_2 t_\beta)}{|M_2|^2 - |\mu|^2} f_{2n} (x_{2L}) - 
\frac{(|\mu|^2 + \mu M_2 t_\beta)}{|M_2|^2 - |\mu|^2} f_{2n} (x_{\mu L}) \right] \nonumber \\
& + \frac{\alpha_Y}{4 \pi} \frac{(m_{LL}^2)_{ij}}{m_L^4} \left[  f_{1n} (x_{1L}) - \frac{(|M_1|^2 + \mu M_1 t_\beta)}{|M_1|^2 - |\mu|^2} f_{2n} (x_{1L}) + 
\frac{(|\mu|^2 + \mu M_1 t_\beta)}{|M_1|^2 - |\mu|^2} f_{2n} (x_{\mu L}) \right] \nonumber \\
& -  \frac{\alpha_Y}{4 \pi} \frac{M_1}{m_{\mu}}  \frac{(m_{LR}^2)_{ji}^*}{m_L^2 - m_R^2} \left[ \frac{1}{m_L^2} f_{3n} (x_{1L}) - \frac{1}{m_R^2} f_{3n} (x_{1R}) \right] \nonumber \\
& -  \frac{\alpha_Y}{4 \pi} \frac{M_1}{m_{\mu}} \frac{(m_{LR}^2 m_{RR}^2)_{ji}^*}{(m_L^2 - m_R^2) m_R^2} \left[ \frac{m_R^2}{m_L^2 - m_R^2}  \left( \frac{1}{m_L^2} f_{3n} (x_{1L}) - \frac{1}{m_R^2} f_{3n} (x_{1R}) \right) +  \frac{2}{m_R^2} f_{2n}(x_{1R}) \right]  \nonumber \\
& +  \frac{\alpha_Y}{4 \pi} \frac{M_1}{m_{\mu}} \frac{(m_{LL}^2 m_{LR}^2)_{ji}^*}{(m_L^2 - m_R^2) m_L^2} \left[ \frac{m_L^2}{m_L^2 - m_R^2}  \left( \frac{1}{m_L^2} f_{3n} (x_{1L}) - \frac{1}{m_R^2} f_{3n} (x_{1R}) \right) +  \frac{2}{m_L^2} f_{2n}(x_{1L}) \right]  \nonumber \\
& +  \frac{\alpha_Y}{2 \pi} \frac{M_1}{m_{\mu}}  \frac{(m_{LL}^2 m_{LR}^2 m_{RR}^2)_{ji}^*}{(m_L^2 - m_R^2)^2 m_L^2 m_R^2} \times \nonumber \\
&  \left[  \frac{m_L^2 m_R^2}{m_L^2 - m_R^2}  \left( \frac{1}{m_L^2} f_{3n} (x_{1L}) - \frac{1}{m_R^2} f_{3n} (x_{1R}) \right) + \frac{m_R^2}{m_L^2} f_{2n}(x_{1L})  + \frac{m_L^2}{m_R^2} f_{2n}(x_{1R})  \right]\,, \\ \nonumber\\
A^{ij(c)}_L & =  \frac{\alpha_2}{4 \pi} \frac{(m_{LL}^2)_{ij}}{m_L^4} \left[  f_{1c} (x_{2L}) + \frac{(|M_2|^2 + \mu M_2 t_\beta)}{|M_2|^2 - |\mu|^2} f_{2c} (x_{2L}) -  
\frac{(|\mu|^2 + \mu M_2 \tan \beta)}{|M_2|^2 - |\mu|^2} f_{2c} (x_{\mu L}) \right]\,,
\end{align}
\begin{align}
A^{ij(n)}_R & = 
\frac{\alpha_Y}{2 \pi} 
\frac{(m_{RR}^2)_{ij}}{m_R^4} \left[ 2 f_{1n} (x_{1R}) +  \frac{(|M_1|^2 + \mu M_1 t_\beta)}{|M_1|^2 - |\mu|^2} f_{2n} (x_{1R}) - 
 \frac{(|\mu|^2 + \mu M_1 t_\beta)}{|M_1|^2 - |\mu|^2} f_{2n} (x_{\mu R}) \right] \nonumber \\
& -  \frac{\alpha_Y}{4 \pi} \frac{M_1}{m_{\mu}}  \frac{(m_{LR}^2)_{ij}}{m_L^2 - m_R^2} \left[ \frac{1}{m_L^2} f_{3n} (x_{1L}) - \frac{1}{m_R^2} f_{3n} (x_{1R}) \right] \nonumber \\
& -  \frac{\alpha_Y}{4 \pi} \frac{M_1}{m_{\mu}} \frac{(m_{LR}^2 m_{RR}^2)_{ij}}{(m_L^2 - m_R^2) m_R^2} \left[ \frac{m_R^2}{m_L^2 - m_R^2}  \left( \frac{1}{m_L^2} f_{3n} (x_{1L}) - \frac{1}{m_R^2} f_{3n} (x_{1R}) \right) +  \frac{2}{m_R^2} f_{2n}(x_{1R}) \right]  \nonumber \\
& +  \frac{\alpha_Y}{4 \pi} \frac{M_1}{m_{\mu}} \frac{(m_{LL}^2 m_{LR}^2)_{ij}}{(m_L^2 - m_R^2) m_L^2} \left[ \frac{m_L^2}{m_L^2 - m_R^2}  \left( \frac{1}{m_L^2} f_{3n} (x_{1L}) - \frac{1}{m_R^2} f_{3n} (x_{1R}) \right) +  \frac{2}{m_L^2} f_{2n}(x_{1L}) \right]  \nonumber \\
& +  \frac{\alpha_Y}{2 \pi} \frac{M_1}{m_{\mu}}  \frac{(m_{LL}^2 m_{LR}^2 m_{RR}^2)_{ij}}{(m_L^2 - m_R^2)^2 m_L^2 m_R^2} \times \nonumber \\
&  \left[  \frac{m_L^2 m_R^2}{m_L^2 - m_R^2}  \left( \frac{1}{m_L^2} f_{3n} (x_{1L}) - \frac{1}{m_R^2} f_{3n} (x_{1R}) \right) + \frac{m_R^2}{m_L^2} f_{2n}(x_{1L})  + \frac{m_L^2}{m_R^2} f_{2n}(x_{1R})  \right]\,.
\end{align}
Here $(m_{LR}^2)_{ii} = m_{l_i}(A_i - \mu^* t_\beta)$, $x_{iA} = |M^2_i|/m^2_A$, $x_{\mu A} = |\mu|^2/m^2_A$ with $i=1,2$ and $A=L,R$.
The explicit expressions for the loop functions are:
\bea
f_{1n}(x)&=&\frac{-17x^3+9x^2+9x-1+6x^2(x+3)\ln x}{24(1-x)^5}\,,\\
f_{2n}(x)&=&\frac{-5x^2+4x+1+2x(x+2)\ln x}{4(1-x)^4}\,,\\
f_{3n}(x)&=&\frac{1+2x\ln x-x^2}{2(1-x)^3}\,,\\
f_{1c}(x)&=&\frac{-x^3-9x^2+9x+1+6x(x+1)\ln x}{6(1-x)^5}\,,\\
f_{2c}(x)&=&\frac{-x^2-4x+5+2(2x+1)\ln x}{2(1-x)^4}\,.
\eea
In the degenerate SUSY limit $M_1 = M_2 = \mu = m_L = m_R = \tilde{m}$ one obtains 
\begin{align}
A^{ij(n)}_L & = - \frac{\alpha_2}{120 \pi} \frac{(m_{LL}^2)_{ij}}{\tilde{m}^4} \left[ \frac{1}{8} + t_\beta \right]  
+ \frac{\alpha_Y}{120 \pi} \frac{(m_{LL}^2)_{ij}}{\tilde{m}^4} \left[  - \frac{5}{8} + t_\beta \right]
+  \frac{\alpha_Y}{48 \pi} \frac{\tilde{m}}{m_{l_i}}  \frac{(m_{LR}^2)_{ji}^*}{\tilde{m}^4}  \nonumber \\
& - \frac{\alpha_Y}{80 \pi} \frac{\tilde{m}}{m_{l_i}} 
\left[
\frac{(m_{LR}^2 m_{RR}^2)_{ji}^*}{\tilde{m}^6} + \frac{(m_{LL}^2 m_{LR}^2)_{ji}^*}{\tilde{m}^6} - 
\frac{2}{3} \frac{(m_{LL}^2 m_{LR}^2 m_{RR}^2)_{ji}^*}{\tilde{m}^8} 
\right]\,,
\\ \nonumber \\
A^{ij(c)}_L & =  \frac{\alpha_2}{40 \pi} \frac{(m_{LL}^2)_{ij}}{\tilde{m}^4} \left[\frac{1}{3} + t_\beta \right]\,, \\ \nonumber \\
A^{(n)}_R & = - \frac{\alpha_Y}{60 \pi} \frac{(m_{RR}^2)_{ij}}{\tilde{m}^4} \left[  \frac{1}{2} + t_\beta \right] +  \frac{\alpha_Y}{48 \pi} \frac{\tilde{m}}{m_{l_i}}  \frac{(m_{LR}^2)_{ij}}{\tilde{m}^4}   -  \frac{\alpha_Y}{80 \pi} \frac{\tilde{m}}{m_{l_i}} \frac{(m_{LR}^2 m_{RR}^2)_{ij}}{\tilde{m}^6 } \nonumber \\
&  -  \frac{\alpha_Y}{80 \pi} \frac{\tilde{m}}{m_{l_i}} \frac{(m_{LL}^2 m_{LR}^2)_{ij}}{\tilde{m}^6}  +  \frac{\alpha_Y}{120 \pi} \frac{\tilde{m}}{m_{l_i}}  \frac{(m_{LL}^2 m_{LR}^2 m_{RR}^2)_{ij}}{ \tilde{m}^8}\,.   
\end{align}

\subsection{Anomalous Magnetic Moments}
The supersymmetric contributions to the anomalous magnetic moment $\Delta a_{l_i}$ come from neutralino and chargino loops 
$\Delta a_{l_i} = \Delta a^{(n)}_{l_i} + \Delta a^{(c)}_{l_i}$. They read:
\begin{align}
\Delta a^{(n)}_{l_i} & =  
\frac{\alpha_2 }{ 8 \pi} \frac{m_{l_i}^2}{m_L^2}   \left[ - f_n^L (x_{2L})+ 2 \frac{|M_2|^2 + {\rm Re}(\mu M_2) \, t_\beta}{|M_2|^2 - |\mu|^2} f_{3n} (x_{2L}) -  
2  \frac{|\mu|^2 + {\rm Re}(\mu M_2) \, t_\beta}{|M_2|^2 - |\mu|^2} f_{3n} (x_{\mu L}) \right]   \nonumber \\
&-   \frac{\alpha_Y}{ 2 \pi}  \frac{m_{l_i}^2}{m_R^2} \left[f_n^L(x_{1R}) - \frac{|M_1|^2 + {\rm Re}(\mu M_1) \, t_\beta}{|M_1|^2 - |\mu|^2} f_{3n} (x_{1R}) + 
\frac{|\mu|^2 + {\rm Re} (\mu M_1) \, t_\beta}{|M_1|^2 - |\mu|^2} f_{3n} (x_{\mu R}) \right] \nonumber \\
& + \frac{\alpha_Y}{ 8 \pi} \frac{m_{l_i}^2}{m_L^2} \left[ - f_n^L (x_{1L})-  2 \frac{|M_1|^2 + {\rm Re}(\mu M_1) \, t_\beta}{|M_1|^2 - |\mu|^2} f_{3n} (x_{1L}) +  
2  \frac{|\mu|^2 + {\rm Re}(\mu M_1) \, t_\beta}{|M_1|^2 - |\mu|^2}  f_{3n} (x_{\mu L})  \right]   \nonumber \\
& +    \frac{\alpha_Y}{ 2 \pi}  \frac{m_{l_i}}{m_L^2 - m_R^2}   {\rm Re}(M_1 m_{LR}^2)_{ii} \left[ \frac{1}{m_L^2} f_{3n}(x_{1L}) - 
\frac{1}{m_R^2} f_{3n}(x_{1R})  \right]  \nonumber \\
& + \frac{\alpha_Y}{ 2 \pi} \frac{m_{l_i}}{m_L^2 - m_R^2} {\rm Re}(M_1 m^2_{LR} m_{RR}^2)_{ii} 
\left[ \frac{\frac{1}{m_L^2} f_{3n}(x_{1L}) - \frac{1}{m_R^2} f_{3n}(x_{1R})}{m_L^2 - m_R^2} + 2 \frac{1}{m_R^4}f_{2n}(x_{1R}) \right]  \nonumber \\
& - \frac{\alpha_Y}{ 2 \pi}  \frac{m_{l_i}}{m_L^2 - m_R^2}  {\rm Re}(M_1 m^2_{LL} m_{LR}^2)_{ii} 
\left[ \frac{\frac{1}{m_L^2} f_{3n}(x_{1L}) - \frac{1}{m_R^2} f_{3n} (x_{1R})}{m_L^2 - m_R^2} +2  \frac{1}{m_L^4} f_{2n}(x_{1L}) \right]  \nonumber \\
& - \frac{\alpha_Y}{\pi}m_{l_i} \frac{{\rm Re}(M_1 m^2_{LL} m^2_{LR} m_{RR}^2)_{ii}}{(m_L^2 - m_R^2)^2}
\left[ \frac{\frac{1}{m_L^2}f_{3n}(x_{1L}) - \frac{1}{m_R^2}f_{3n}(x_{1R})}{m_L^2 - m_R^2} + \frac{f_{2n}(x_{1L})}{m_L^4} +  \frac{f_{2n} (x_{1R})}{m_R^4} \right]\,,   
\\
\Delta a^{(c)}_{l_i} & =\frac{\alpha_2}{ 4 \pi} \frac{m_{l_i}^2}{m_L^2} \left[ f^L_c (x_{2L}) - \frac{|M_2|^2 + {\rm Re}(\mu M_2) \, t_\beta}{|M_2|^2-|\mu|^2} f_c^{LR} (x_{2L}) +
\frac{|\mu|^2 + {\rm Re}(\mu M_2) \,  t_\beta}{|M_2|^2-|\mu|^2} f_c^{LR} (x_{\mu L}) \right]\,. 
\end{align}
Here we introduced the additional loop functions:
\bea
f^L_n (x)&=&\frac{1-6x+3x^2+2x^3-6x^2 \log x}{6 (1-x)^4}\,, \\
f^L_c (x)&=&\frac{2+3x-6x^2+x^3 + 6 x \log x}{6 (1-x)^4}\,, \\
f^{LR}_c (x)&=&\frac{-3 + 4 x - x^2- 2  \log x}{ (1-x)^3} \,.
\eea
In the degenerate SUSY limit $M_1 = M_2 = \mu = m_L = m_R = \tilde{m}$ one obtains 

\begin{align}
\Delta a^{(n)}_{l_i} & =   
- \frac{\alpha_2 }{ 48 \pi} \frac{m_{l_i}^2}{\tilde{m}^2}   \left[ - \frac{1}{2} +  \frac{{\rm Re} (\mu M_2)}{\tilde{m}^2} t_\beta \right]
-\frac{\alpha_Y}{48 \pi}  \frac{m_{l_i}^2}{\tilde{m}^2}   
\left[
\left( \frac{3}{2} + \frac{{\rm Re} (\mu M_1)}{\tilde{m}^2} t_\beta \right)  
+\frac{2{\rm Re} (M_1 m_{LR}^2)_{ii}}{m_{l_i}\tilde{m}^2} \right]
\nonumber \\ + 
&\frac{\alpha_Y}{ 40 \pi}  \frac{m_{l_i}}{\tilde{m}^2} 
\left[
\frac{{\rm Re} (M_1 m^2_{LR} m_{RR}^2)_{ii}}{\tilde{m}^4} + \frac{{\rm Re} (M_1 m^2_{LL} m_{LR}^2)_{ii}}{\tilde{m}^4} -    
\frac{2}{3} \frac{{\rm Re} (M_1 m^2_{LL} m^2_{LR} m_{RR}^2)_{ii}}{\tilde{m}^6}
\right]\,,\\
\Delta a^{(c)}_{l_i} & =    \frac{\alpha_2}{ 8 \pi} \frac{m_{l_i}^2}{\tilde{m}^2} \left[ - \frac{1}{6} + \frac{{\rm Re} (\mu M_2)}{\tilde{m}^2} t_\beta \right]\,. 
\end{align}
\subsection{Electric Dipole Moments}

The supersymmetric contributions to the Electric Dipole Moment $d_i$ come from neutralino and chargino loops  $d_i = d^{(n)}_i + d^{(c)}_i$. They read:
\begin{align}
\frac{d^{(n)}_i}{e} & = 
\frac{\alpha_2 }{ 8 \pi} \frac{m_{l_i}}{m_L^2} \frac{ {\rm Im}(\mu M_2)}{|M_2|^2 - |\mu|^2} t_\beta \left[ f_{3n} (x_{2L}) -  f_{3n}(x_{\mu L}) \right] \nonumber \\
& +\frac{\alpha_Y}{ 4 \pi} \frac{m_{l_i}}{m_R^2} \frac{{\rm Im}(\mu M_1)}{|M_1|^2 - |\mu|^2} t_\beta \left[ f_{3n}(x_{1R}) - f_{3n} (x_{\mu R}) \right] \nonumber \\
& - \frac{\alpha_Y}{ 8 \pi} \frac{m_{l_i}}{m_L^2} \frac{{\rm Im} (\mu M_1)}{|M_1|^2 - |\mu|^2} t_\beta \left[ f_{3n}(x_{1L}) - f_{3n}(x_{\mu L}) \right] \nonumber \\
& + \frac{\alpha_Y}{ 4 \pi} \frac{{\rm Im} (M_1 m_{LR}^2)_{ii}}{m_L^2 - m_R^2} \left[ \frac{1}{m_L^2} f_{3n}(x_{1L}) - \frac{1}{m_R^2} f_{3n}(x_{1R})  \right]  \nonumber \\
& + \frac{\alpha_Y}{ 4 \pi} \frac{{\rm Im}(M_1 m^2_{LR} m_{RR}^2)_{ii}}{m_L^2 - m_R^2} \left[ \frac{\frac{1}{m_L^2} f_{3n}(x_{1L}) - \frac{1}{m_R^2} f_{3n}(x_{1R})}{m_L^2 - m_R^2} + 2 \frac{1}{m_R^4} f_{2n} (x_{1R}) \right]  \nonumber \\
& -    \frac{\alpha_Y}{ 4 \pi}  \frac{{\rm Im} (M_1 m^2_{LL} m_{LR}^2)_{ii}}{m_L^2 - m_R^2} 
\left[ \frac{\frac{1}{m_L^2} f_{3n}(x_{1L}) - \frac{1}{m_R^2} f_{3n}(x_{1R})}{m_L^2 - m_R^2} + 2 \frac{1}{m_L^4} f_{2n} (x_{1L}) \right]  \nonumber \\
& -    \frac{\alpha_Y}{ 2 \pi} \frac{ {\rm Im} (M_1 m^2_{LL} m^2_{LR} m_{RR}^2)_{ii}}{(m_L^2 - m_R^2)^2} \times \nonumber \\
& \left[ \frac{\frac{1}{m_L^2}f_{3n}(x_{1L}) - \frac{1}{m_R^2} f_{3n}(x_{1R})}{m_L^2 - m_R^2} + \frac{1}{m_L^4} f_{2n}(x_{1L}) + \frac{1}{m_R^4} f_{2n}(x_{1R}) \right]\,,   \\
\frac{d^{(c)}_i}{e} & =   
- \frac{\alpha_2}{8\pi} \frac{m_{l_i}}{m_L^2} \frac{{\rm Im} (\mu M_2)}{|M_2|^2-|\mu|^2} t_\beta \left[ f_c^{LR} (x_{2L}) -  f_c^{LR} (x_{\mu L})  \right]\,.
\end{align}
In the degenerate SUSY limit $M_1 = M_2 = \mu = m_L = m_R = \tilde{m}$ one obtains 

\begin{align}
\frac{d^{(n)}_i}{e} & = 
- \frac{\alpha_2 }{ 96 \pi} \frac{m_{l_i}}{\tilde{m}^2}  \frac{ {\rm Im} (\mu M_2)}{\tilde{m}^2}  t_\beta
- \frac{\alpha_Y}{ 96 \pi} \frac{m_{l_i}}{\tilde{m}^2} \frac{ {\rm Im} (\mu M_1)}{\tilde{m}^2} t_\beta    
- \frac{\alpha_Y}{ 120 \pi}   \frac{{\rm Im}  (M_1 m^2_{LL} m^2_{LR} m_{RR}^2)_{ii}}{\tilde{m}^8}
\nonumber \\
& - \frac{\alpha_Y}{ 48 \pi} \frac{{\rm Im} (M_1 m_{LR}^2)_{ii}}{\tilde{m}^4} + \frac{\alpha_Y}{ 80 \pi} 
\frac{{\rm Im}  (M_1 m^2_{LR} m_{RR}^2)_{ii}}{\tilde{m}^6} +    
\frac{\alpha_Y}{ 80 \pi}   \frac{{\rm Im}  (M_1 m^2_{LL} m_{LR}^2)_{ii}}{\tilde{m}^6}\,,
\\ \nonumber\\
\frac{d^{(c)}_i}{e} & =    \frac{\alpha_2}{ 16 \pi} \frac{m_{l_i}}{\tilde{m}^2} \frac{ {\rm Im} (\mu M_2)}{\tilde{m}^2}  t_\beta\,.
\end{align}

\end{document}